\documentclass[twocolumn,english]{IEEEtran}
\usepackage[T1]{fontenc}
\usepackage{color}
\usepackage{babel}
\usepackage{array}
\usepackage{verbatim}
\usepackage{float}
\usepackage{textcomp}
\usepackage{multirow}
\usepackage{amsthm}
\usepackage{graphicx}
\usepackage[unicode=true,
 bookmarks=true,bookmarksnumbered=true,bookmarksopen=true,bookmarksopenlevel=1,
 breaklinks=false,pdfborder={0 0 0},backref=false,colorlinks=false]
 {hyperref}
\hypersetup{pdftitle={Your Title},
 pdfauthor={Your Name},
 pdfpagelayout=OneColumn, pdfnewwindow=true, pdfstartview=XYZ, plainpages=false}

\makeatletter

\providecommand{\tabularnewline}{\\}

\theoremstyle{plain}
\newtheorem{thm}{\protect\theoremname}
\theoremstyle{plain}
\newtheorem{prop}[thm]{\protect\propositionname}

\ifCLASSOPTIONcompsoc
\usepackage[caption=false,font=normalsize,labelfont=sf,textfont=sf]{subfig}
\else
\usepackage[caption=false,font=footnotesize]{subfig}
\fi

\usepackage{lipsum}
\newcommand\blfootnote[1]{%
  \begingroup
  \renewcommand\thefootnote{}\footnote{#1}%
  \addtocounter{footnote}{-1}%
  \endgroup
}

\usepackage{color,xcolor}
\definecolor{red}{RGB}{255,0,0}

\usepackage{amssymb}
\usepackage{amsmath}
\usepackage{needspace}

\makeatother

\providecommand{\propositionname}{Proposition}
\providecommand{\theoremname}{Theorem}
\graphicspath{ {/Figures/} }
\begin{document}

\title{Using the Best Linear Approximation With Varying Excitation Signals
for Nonlinear System Characterization}

\author{Alireza~Fakhrizadeh~Esfahani,~\IEEEmembership{Student Member IEEE,}
Johan~Schoukens,~\IEEEmembership{Fellow, IEEE,} and~Laurent~Vanbeylen,~\IEEEmembership{Member,~IEEE}}

\maketitle

\IEEEspecialpapernotice{}

\IEEEaftertitletext{}

\medskip{}

\begin{abstract}
Block oriented model structure detection is quite desirable since
it helps to imagine the system with real physical elements. In this
work we explore experimental methods to detect the internal structure
of the system, using a black box approach. Two different strategies
are compared and the best combination of these is introduced. The
methods are applied on two real systems with a static nonlinear block
in the feedback path. The main goal is to excite the system in a way
that reduces the total distortion in the measured frequency response
functions to have more precise measurements and more reliable decision
about the structure of the system.%
\end{abstract}

\section{Introduction}

\IEEEPARstart{F}{requency} response function (FRF) measurement,
is very interesting for modelling and identification. FRF analysis
is used not only for modelling but also it can be used to get insight
in the internal structure of the system under test. %
Since nonlinear (NL) identification usually is a very time consuming
procedure, it is highly desirable to carefully select an appropriate
model structure prior to the actual identification. In this article,
we study block-oriented systems \cite{Billings}, and \cite{Fouad},
consisting of interconnected dynamic linear blocks and static nonlinear
blocks. In \cite{Billings} different block-oriented structure systems,
such as, Wiener-Hammerstein, parallel Wiener-Hammerstein and nonlinear
unity feedback systems are studied in the time domain. These structures
are analyzed by using spectral analysis of Volterra series expansion
of the block structured non-linear system in \cite{Billings3}. A
complete review of these methods can be found in \cite{Haber1}, and
\cite{Fouad}. 
\blfootnote{\hspace{-4mm}\noindent\rule{9cm}{0.4pt} \\ \hspace*{5mm} The authors thank Johan Pattyn for the design and realization of the setup in \ref{Block-diagram-of}$(b)$. \\ \hspace*{5mm} This work was supported in part by the Fund for Scientific Research (FWO-Vlaanderen), by the Flemish Government (Methusalem), the Belgian Government through the Inter university Poles of Attraction (IAP VII) Program, and by the ERC advanced grant SNLSID, under contract 320378.\vspace{2mm} \scriptsize{Using the Best Linear Approximation With Varying Excitation Signals for Nonlinear System Characterization\\
Alireza Fakhrizadeh Esfahani; Johan Schoukens; Laurent Vanbeylen\\
IEEE Transactions on Instrumentation and Measurement\\
Year: 2016, Volume: 65, Issue: 5 Pages: 1271 - 1280\\
DOI: 10.1109/TIM.2015.2504079\\
https://ieeexplore.ieee.org/document/7373634/ \\
\copyright IEEE 2016}}

Here the internal structure of block oriented systems is explored,
for which already a large number of possible internal structures can
be present. %
In this work the focus is on the detection of the nonlinear structure
by using the Best Linear Approximation (BLA), such as, the presence
of NL feedback. The FRF of the BLA \cite{Rik1} is measured for different
excitation signal's properties. In Billings and Fakhouri \cite{Billings2},
it is proposed to use nonzero mean Gaussian distributed signals in
combination with a correlation analysis, for recognizing a nonlinear
system with a unity feedback structure. In \cite{Lieve}, the standard
deviation (STD) and bandwidth of the excitation signals are changed
to study the internal structure of the system by using the BLA analysis.\textcolor{black}{{}
In \cite{khodam} (This paper is the extension of the article, which
authors presented in the I\texttwosuperior{}MTC conference in 2015),
}in contrast to \cite{Lieve}, amplitude (STD value) level, and/or
DC level (mean value, offset or operating point) of a full-band random
phase multisine within frequency band of interest are varied to analyze
the block oriented nonlinear structure\textcolor{black}{. In this
study, the idea of varying DC and STDs, as proposed in \cite{khodam}
is extended to find an optimal selection of DC and STD levels, to
have a minimum amount of distortions and therefore a set of high-quality
BLA measurements (which, on turn, can serve to get insight in the
nonlinear system). The proposed method is based on the Central Composite
Design (CCD) method \cite{Heikki}. The novelty of the article is
the extraction of an eigen-path using CCD, which gives the best}%
\footnote{\textcolor{black}{The strategy is optimal in the sense that - in a
local approximation in the DC/STD space - the magnitudes of the total
distortions on the BLA are jointly minimized.}%
}\textcolor{black}{{} strategy for varying the DC, and STD levels of
the input signal.}

\textcolor{black}{Multisines are used for exciting systems under test.
Multisines are in the class of Gaussian signals by increasing the
number of frequencies lines asymptotically \cite{Rik1}. This periodic
signal gives the same BLA as a filtered Gaussian (a colored Gaussian
process) exc}itation with the same bandwidth.%

Simultaneously with the measurement of the BLA, a nonparametric analysis
of the nonlinear distortions is done. On the basis of these results
the potential of both properties is discussed. A method is proposed
for which both properties (DC and STD levels) are varied. By using
this method, FRFs have less distortion, and the structure detection
and furthermore any post-processing calculation will be more reliable.

The internal nonlinear structure of the system can be identified,
by changing the excitation's characteristic values, the DC and STD
levels. %
Information about the nonlinear structure of the system can be retrieved
by finding the BLA, with different experiments through varying the
excitation signal's DC and STD values. 

\textcolor{black}{}%
\textcolor{black}{}%
\textcolor{black}{In Section \ref{sec:Bussgang's-Theorem-For}, Bussgang's
theorem \cite{Bussgang}, (providing information about the BLA of
static systems), is proved to be valid for an excitation contains
DC value.} It is shown, that varying the DC level of the excitation
signal has less nonlinear distortion contributions at the BLA. So
it is a good idea to vary the DC level, to retrieve information about
the nonlinear structure of the system through variations of the BLA.

By calculating and fitting a parametric model on the FRF, we can follow
the behavior of the poles and zeros and choose the best internal structure\textcolor{black}{{}
\cite{Johan1}}.

The main contributions of this work are:
\begin{itemize}
\item Extension of Bussgang's theorem to include nonzero mean Gaussian inputs.
\item Structure detection for block oriented models, through varying the
input properties. %

\item Analysis of the pole-zero behavior of different internal structures.
\item Finding the best strategy for varying the DC and STD level of the
signal to measure different high quality BLAs that can be used to
detect the block oriented structure of the system.
\end{itemize}
A proposition to Bussgang's theorem for non-zero mean excitations
is presented in Section \ref{sec:Bussgang's-Theorem-For}. In Section
\ref{sec:Block-oriented-nonlinear}%
, block oriented models are introduced. In Section \ref{sec:The-Best-Linear},
the concept of the best linear approximation is reviewed. %
{} %
The structure detection method is explained in Section \ref{sec:Structure-Detection}%
. In Section \ref{sec:Optimizing,-signal's-characteris} an experimental
method for selecting the best DC and STD levels of the signal is introduced.
Systems under study and measurement results are given in Section \ref{sec:Systems-under-study}
and \ref{sec:Results}%
. Section \ref{sec:Conclusions}%
, concludes the paper. %
\section{Bussgang's Theorem For Non-Zero Mean Excitations\label{sec:Bussgang's-Theorem-For}}

Structure detection, based on the BLA of block oriented models, needs
to study the BLA of the static nonlinearity (SNL), a memoryless function
block. It can be done through the Bussgang's theorem, which simply
stated, means that a SNL system excited with Gaussian signals has
a constant BLA (static, frequency-independent). However, Bussgang
did not provide the proof in case the input has non-zero mean. Here,
it is needed to analyze the BLAs at different DC levels (non-zero
mean input).

According to Bussgang's theorem, \cite{Bussgang} if the input signal
of a SNL system, is a stationary (colored) zero mean Gaussian process,
the BLA of a SNL block is a constant. The static nonlinearity should
fulfill the following property:

\begin{equation}
\int_{-\infty}^{\infty}xg_x(x)e^{-\frac{x^2}{2}}dx< \infty
\end{equation}This condition covers a large class of functions such as, functions
with a finite number of discontinuities, and functions with a finite
number of discontinuous derivatives.

\begin{figure}[tbh]
\vspace{-7mm}

\begin{centering}
\includegraphics[scale=0.25]{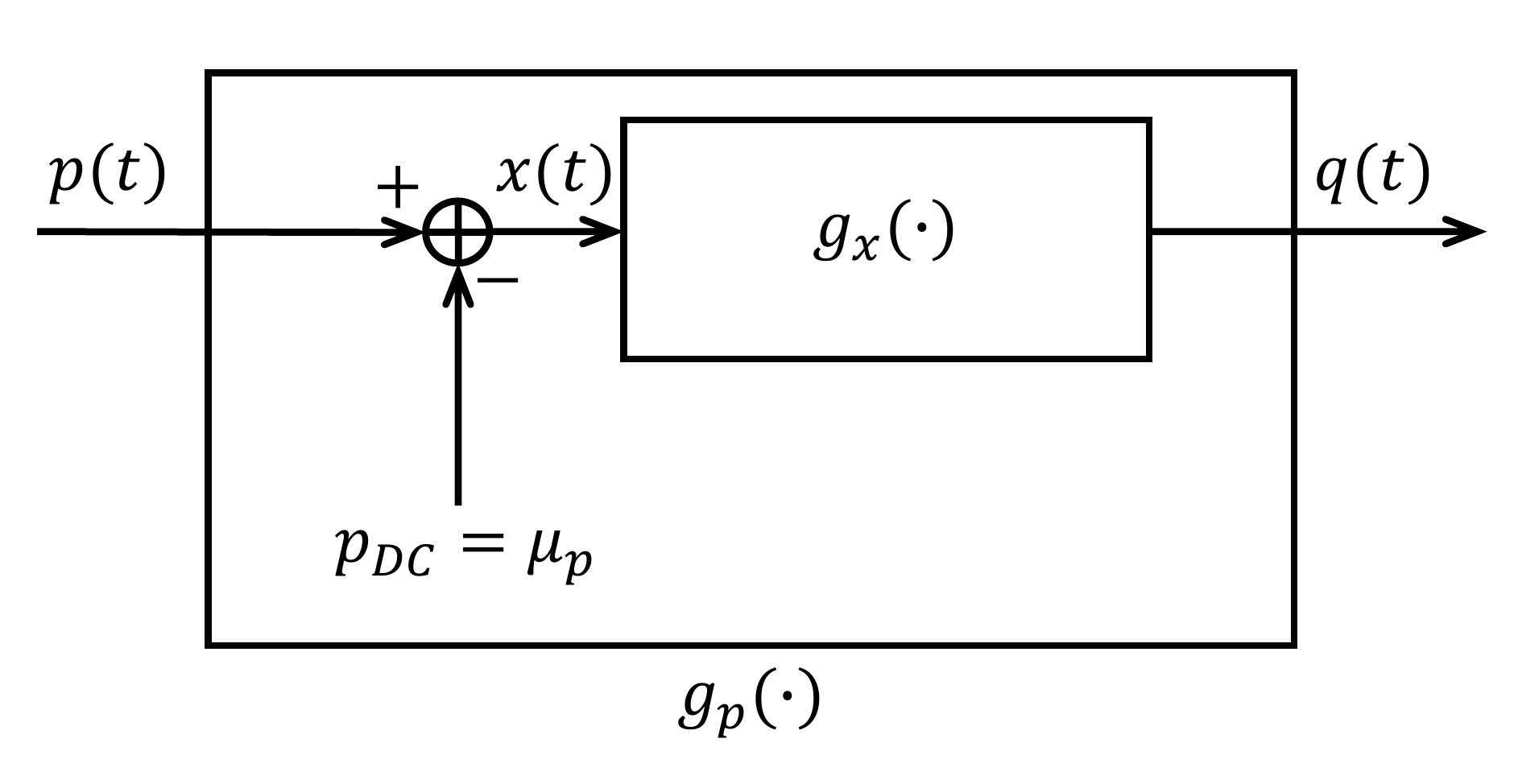}
\par\end{centering}

\vspace{-5mm}

\caption{Block diagram of a SNL system with nonzero mean input.\label{Bussgang}}

\vspace{-3mm}
\end{figure}

Here we will show, that Bussgang's theorem is also valid for non-zero
mean input.
\begin{prop}
\label{prop:Assume--be}Assume $p(t)$ to be a stationary (colored)
nonzero mean Gaussian process, and $q(t)=g_p(p(t))$, then

\vspace{-5mm}
\begin{eqnarray}
R_{pq}(\tau)=kR_{pp}(\tau)
\end{eqnarray}here, $R_{uv}$%
{} and $R_{uu}$%
{} are the cross- and auto-covariance respectively.\end{prop}
\begin{IEEEproof}
It is assumed, that the input process $p(t)$ is a Gaussian process,
with a non-zero mean $\mathbb{E}\left \{ p_{t} \right \}=\mu_p=p_{DC}$
\vspace{-3mm}
\begin{equation}
p(t)=\mu_p+x(t)
\end{equation}where $\mu_p$ is the mean value of $p(t)$ and $x(t)$ is a stationary
zero-mean, Gaussian process (See Fig. \ref{Bussgang}). It can be
remarked that the SNL $g_p(\bullet)$ can be redefined in terms of
an offset and a function $g_x(\bullet)$:

\vspace{-2mm}
\begin{equation}
g_p(p)=g_x(p-\mu_p)
\end{equation}
\vspace{-6mm}or equivalently:

\begin{equation}
g_x(x)=g_p(x+\mu_p)
\end{equation}

\vspace{-2mm}
Since $\mathbf{E} \{ x(t) \} = 0$, Bussgang's theorem applies to
$g_x(\bullet)$ \cite{Bussgang}:

\vspace{1mm}
\begin{equation}
R_{xq}(\tau)= k R_{xx}(\tau)
\label{eq:Crosscorr}
\end{equation}
\vspace{-2mm}

Using the equations corresponding to the block diagram in Fig. \ref{Bussgang},
a result similar to (\ref{eq:Crosscorr}) can be obtained in terms
of the signal $p$:

\vspace{-4mm}
\begin{equation}
R_{pq}(\tau)=kR_{pp}(\tau)
\end{equation}

This fact is the simple consequence of

\vspace{-5mm}
\begin{eqnarray*}
R_{pq}(\tau)&=&\mathbb{E} \left \{ ( p(t+\tau)-\mathbb{E} \left \{ p(t+\tau) \right \} )(q(t)-\mathbb{E} \left \{ q(t) \right \}) \right \}
\\
&=&\mathbb{E} \{ (x(t+\tau)+\mu_{p}-\mathbb{E} \{ x(t+\tau) + \mu_{p } \})
\\
& &(q(t)-\mathbb{E} \{ q(t) \}) \}
\\
&=&\mathbb{E} \{ (x(t+\tau) - \mathbb{E}\{ x(t+\tau) \})(q(t)-\mathbb{E} \{ q(t) \}) \}
\end{eqnarray*}

\vspace{-6mm}
\begin{equation}
\hspace{-46.9mm}=R_{xq}(\tau)
\end{equation}

and

\vspace{-5mm}
\begin{eqnarray*}
R_{pp}(\tau)&=&\mathbb{E} \{ (p(t+\tau)-\mathbb{E} \{ p(t+\tau) \}) (p(t)-\mathbb{E} \{ p(t) \}) \}
\\
&=&\mathbb{E} \{ (p(t+\tau)- \mu_p )(p(t)-\mu_p) \}
\\
&=&\mathbb{E} \{ x(t+\tau) x(t) \}
\\
&=&\mathbb{E} \{ ( x(t+\tau) - \mathbb{E} \{ x(t+\tau) \})(x(t)-\mathbb{E} \{ x(t) \})\}
\end{eqnarray*}

\vspace{-6mm}
\begin{equation}
\hspace{-45.9mm}=R_{xx}(\tau)
\end{equation}

\end{IEEEproof}

In this work multisine excitations are used. Multisines, are among
the class of Gaussian signals, (as number of the excited frequencies
approaches to infinity). Gaussian signals are separable, for separable
processes, and Bussgang's theorem, also known as the invariance property
holds\textcolor{black}{{} \cite{Billings}}, \textcolor{black}{\cite{Nuttall}}.

From proposition \ref{prop:Assume--be}, it follows that Bussgang's
theorem is valid also for nonzero mean excitations. So the best linear
approximation of an SNL system, even with nonzero mean inputs is a
constant $k$ (see also (\ref{eq:Buss1}) and (\ref{eq:Buss2})) depending
on the mean (DC) and the variance (STD\texttwosuperior{}) of $p(t)$%
. %
{} This fact will be used in the sequel to detect the internal structure
of block oriented models (see Section \ref{sec:Structure-Detection}).
{}

\section{Block oriented nonlinear models\label{sec:Block-oriented-nonlinear}}

\begin{figure}[tbh]
\vspace{-0.5mm}

\includegraphics[clip=1cm 0bp 890bp 726bp,scale=0.25]{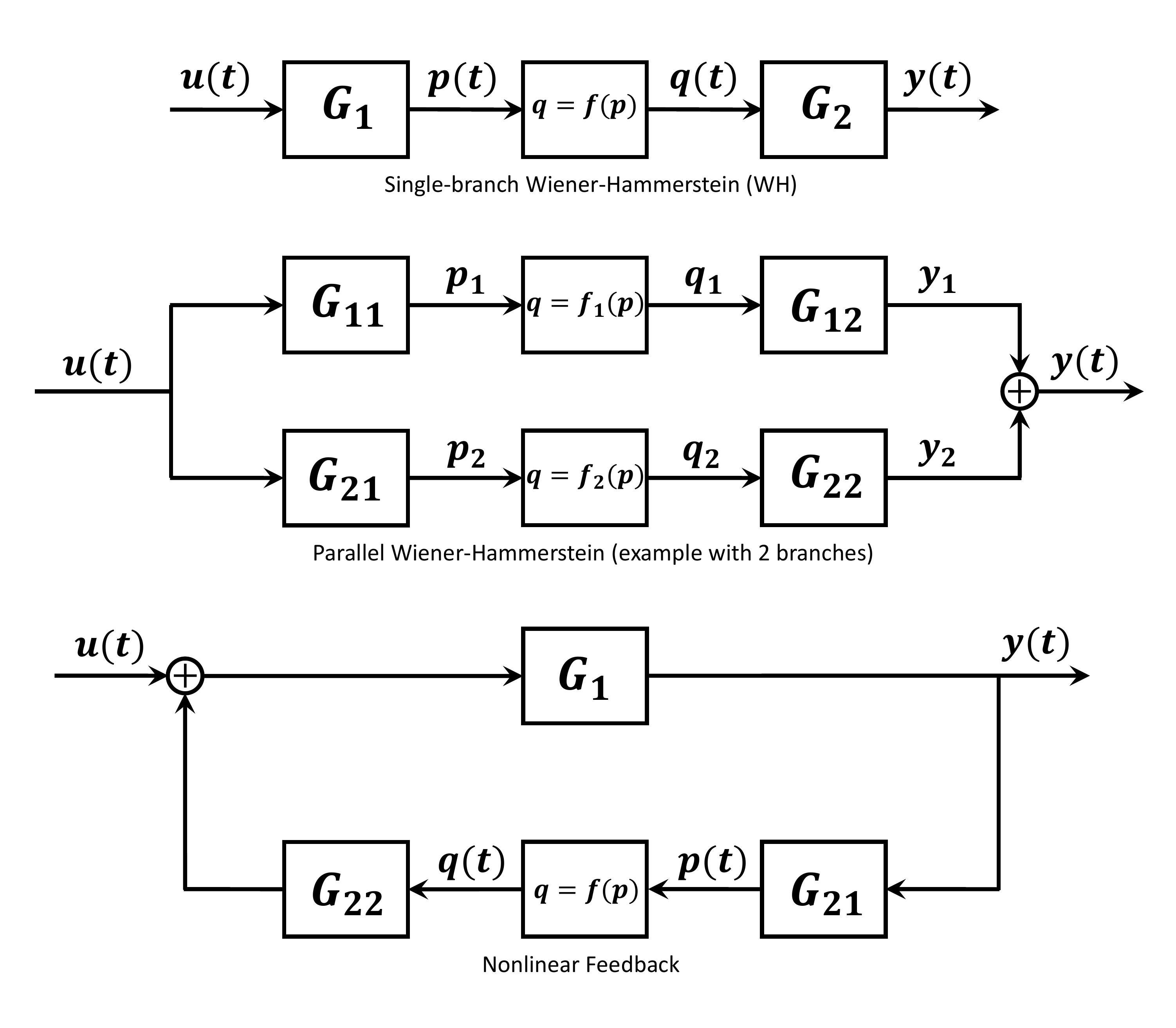}

\vspace{-6mm}

\caption{A few examples of block oriented nonlinear structures.\label{Block_oriented}}

\vspace{-7mm}
\end{figure}

Block oriented nonlinear models can be defined as an interconnection
of static nonlinear (SNL) and linear time invariant blocks \textcolor{black}{\cite{Billings}},
\cite{Fouad}. Three types of block oriented systems that are considered
in this paper are:

\begin{itemize}
\item A Wiener-Hammerstein (WH) system:
\end{itemize}
Is defined as the cascade connection of two linear dynamic systems
and a static nonlinearity in between.
\begin{itemize}
\item A parallel WH system:
\end{itemize}
Parallel branches of different WH systems
\begin{itemize}
\item A nonlinear feedback system:
\end{itemize}
A branch of a WH system at the feedback and a linear time invariant
block at the feedforward path. The WH system can be located either
in the feedforward or in the feedback path \cite{Johan2}. %
\textcolor{black}{In Fig. \ref{Block_oriented}, the three different
bl}ock oriented systems, which are considered in this paper, are shown.

\vspace{-1mm}

\section{The Best Linear Approximation\label{sec:The-Best-Linear}}

The best linear approximation (BLA) is a tool to measure the frequency
response function\textcolor{black}{{} \cite{Rik1}, \cite{Bendat}},\textcolor{black}{{}
\cite{Johan4}} of a nonlinear dynamic system (assuming that the response
to a periodic input is periodic with the same period as the input).
Here the BLA is used to analyze the block oriented internal structure
of the system. In this section the BLA is defined in the first part,
and next the estimation method is explained.

\vspace{-2mm}

\subsection{Definition}

The best linear approximation is defined as \cite{Rik1,Johan4,Enqvist2}

\begin{equation}
G_{BLA}(j\omega)=\underset{G(j\omega)}{\arg\min\:}\mathbb{E}_{u}\left\{ \left|Y(j\omega)-G(j\omega)U(j\omega)\right|^{2}\right\}
\end{equation}where $Y(j\omega)$, $G(j\omega)$, and $U(j\omega)$ are the real
output of the system, the model, and the input of the system respectively.
$G_{BLA}(j\omega)$ is the BLA of the system. $\mathbb{E}_{u}$ denotes
the mathematical expectation with respect to the random input.

\begin{figure}[tbh]
\vspace{0mm}

\includegraphics[clip=20bp 30bp 640bp 160bp,scale=0.38]{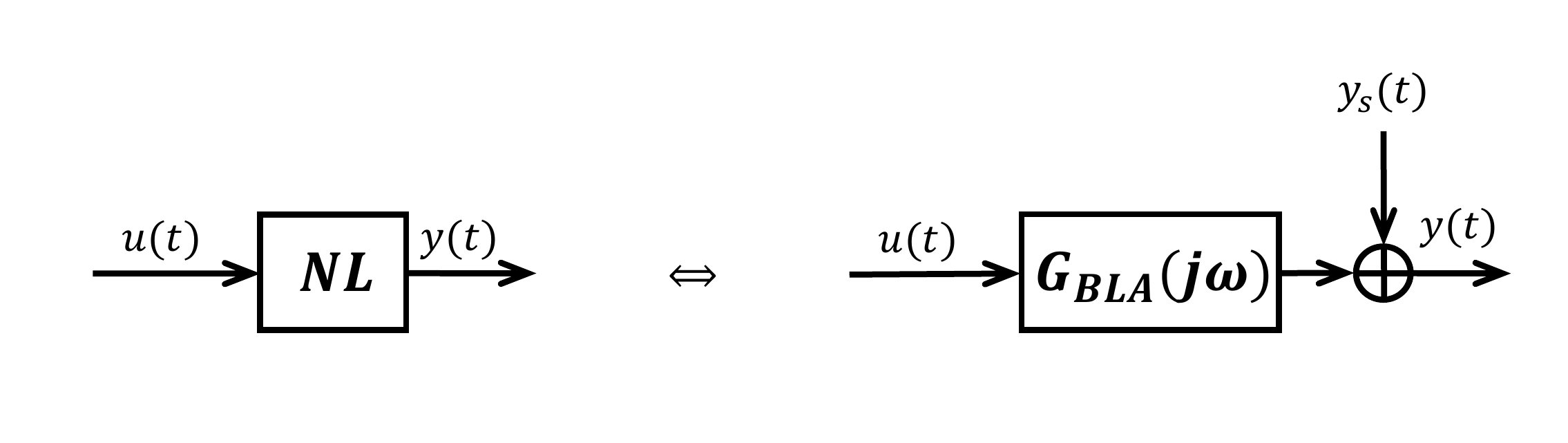}

\vspace{-1mm}

\caption{Analogy between the NL system and the BLA. \label{BLA_Block}}

\vspace{-1mm}
\end{figure}

It can be assumed that a nonlinear dynamic system, is replaced by
the BLA plus a nonlinear distortion term $y_s(t)$ (see Fig. \ref{BLA_Block}).%
\textcolor{black}{{} $y_s(t)$ is uncorrelated with $u(t)$, but not
independent (for a detailed discussion see Sec. 3.4 of \cite{Rik1}
or \cite{Rik2,Johan7})}%
\textcolor{black}{. The following equations hold (see Section \ref{sec:Bussgang's-Theorem-For})
in the time and frequency domain, respectively:}

\vspace{-6mm}
\begin{eqnarray}
R_{yu}&=&g_{BLA} \star R_{uu},
\label{eq:Buss1}
\\
S_{yu}&=&G_{BLA} S_{uu}.
\label{eq:Buss2}
\end{eqnarray}

\vspace{-1mm}
where $R_{yu}$, and $R_{uu}$ are input-output cross correlation,
and input auto-correlation, respectively. $\star$ stands for the
convolution operation, and $g_{BLA}$ is the BLA in the time domain
(the inverse Fourier transform of $G_{BLA}$). $S_{yu}$, and $S_{uu}$
are input-output cross power, and input auto-power spectra, respectively.

\vspace{-3mm}

The BLA depends on the class of inputs to which $u$ belongs. Here,
the class of band-limited Gaussian noise signals, with a certain mean
value (DC) and standard deviation (STD), are used.

\vspace{-5mm}

\subsection{Estimation}

\vspace{0mm}

To estimate the BLA easily, it is proposed to excite the system with
Gaussian distributed excitations (see Section \ref{sec:Bussgang's-Theorem-For})
for example, random phase multisine signals. %
More precisely, with random phase multisines \cite{Rik1}, the calculated
BLA has less uncertainty, compared with random excitations \cite{Rik1},
\textcolor{black}{\cite{Johan3}}.\textcolor{black}{{} Two methods to
analyze the NL distortion with the BLA spectrum, are called the robust
method and the fast method \cite{Rik1}, \cite{Johan3}.}\textcolor{red}{{}
}%

These methods allow one to separate the noise %
and%
{} the nonlinear distortion $y_s(t)$. Hence from the multisine measurement,
the FRF of the BLA, the noise, and the n\textcolor{black}{onlinear
contributions can be quantified non-parametrically in the frequency
domain }%
\textcolor{black}{\cite{Rik1}, \cite{Johan3}. In this work the robust
method is adopted \cite{Johan3,Rik1}.}

\textcolor{black}{The total distortion $\hat{\sigma}^2_{\hat{G}_{BLA}}$,
includes the stochastic nonlinear distortion $\hat{\sigma}^2_{\hat{G}_s}$
on the BLA, and the noise contribution $\hat{\sigma}^2_{\hat{G}_{BLA,n}}$.
By }exciting the system with a pe\textcolor{black}{riodic excitation,
the user is able to estimate the noise contribution $\hat{\sigma}^2_{\hat{G}_{BLA,n}}$.
By subtracting the noise contribution from the total distortion, the
nonlinear distortion is calculated. The mentioned quantities are calculated
as follows:}

\vspace{-10mm}

\vspace{8mm}
\begin{equation}\hspace{-19.17mm}
\hat{U}^{[m]}(j\omega_k) = \frac{1}{P}\sum_{p=1}^{P}U^{[m,p]}(j\omega_k)
\end{equation}
\begin{equation}\hspace{-19.17mm}\vspace{-3mm}
\hat{Y}^{[m]}(j\omega_k) = \frac{1}{P}\sum_{p=1}^{P}Y^{[m,p]}(j\omega_k)
\end{equation}
\begin{equation}\hspace{9.89mm}\vspace{-3mm}
\hat{\sigma}^2_{\hat{U}^{[m]}}(j\omega_k) =\sum_{p=1}^{P}\frac{\left |U^{[m,p]}(j\omega_k)-\hat{U}^{[m]}(j\omega_k)\right |^2}{P(P-1)}
\end{equation}
\begin{equation}\hspace{9.81mm}\vspace{-3mm}
\hat{\sigma}^2_{\hat{Y}^{[m]}}(j\omega_k) =\sum_{p=1}^{P}\frac{\left |Y^{[m,p]}(j\omega_k)-\hat{Y}^{[m]}(j\omega_k)\right |^2}{P(P-1)}
\end{equation}

\vspace{0mm}

\begin{equation}
\begin{matrix}
\hspace{-4.481cm}
\vspace{0mm}\hat{\sigma}^2_{\hat{Y}^{[m]}\hat{U}^{[m]}}(j\omega_k) = \sum_{p=1}^{P}\vspace{1mm}\\
\frac{(Y^{[m,p]}(j\omega_k)-\hat{Y}^{[m]}(j\omega_k))(\overline{U^{[m,p]}(j\omega_k)-\hat{U}^{[m]}(j\omega_k))}}{P(P-1)}\\
\end{matrix}
\vspace{7mm}
\end{equation}

\vspace{0mm}

\textcolor{black}{\vspace{-10mm}
\begin{equation}\hspace{-30.26mm}
\hspace{-0.78cm}\hat{G}^{[m]}(j\omega_k) = \frac{\hat{Y}^{[m]}(j\omega_k)}{\hat{U}^{[m]}(j\omega_k)}
\end{equation}
\begin{equation}\hspace{-25.88mm}
\hspace{-0.78cm}\hat{G}_{BLA}(j\omega_k)=\sum_{m=1}^{M}\frac{\hat{G}^{[m]}(j\omega_k)}{M}
\end{equation}
\begin{equation}\hspace{-0.87mm}
\hspace{-0.78cm}\hat{\sigma}^2_{\hat{G}_{BLA}}(j\omega_k) =\sum_{m=1}^{M}\frac{\left |G^{[m]}(j\omega_k)-\hat{G}_{BLA}(j\omega_k)\right |^2}{M(M-1)}
\end{equation}
\begin{equation}\hspace{-18.12mm}
\hspace{-0.78cm}\hat{\sigma}^2_{\hat{U},n}(j\omega_k) = \frac{1}{M}\sum_{m=1}^{M}\hat{\sigma}^2_{\hat{U}^{[m]}}(j\omega_k)
\end{equation}
\begin{equation}\hspace{-18.12mm}
\hspace{-0.78cm}\hat{\sigma}^2_{\hat{Y},n}(j\omega_k) = \frac{1}{M}\sum_{m=1}^{M}\hat{\sigma}^2_{\hat{Y}^{[m]}}(j\omega_k)
\end{equation}
\begin{equation}\hspace{-14.37mm}
\hspace{-0.78cm}\hat{\sigma}^2_{\hat{Y}\hat{U},n}(j\omega_k) = \frac{1}{M}\sum_{m=1}^{M}\hat{\sigma}^2_{\hat{Y}^{[m]}\hat{U}^{[m]}}(j\omega_k)
\end{equation}
\begin{equation}\hspace{-9.59mm}
\begin{matrix}
\vspace{2mm}
\hspace{-3.225cm}\hat{S}_{U_{0}U_{0}}(j\omega_k)=\frac{1}{M}\sum_{m=1}^{M}\\
(\left |\hat{U}^{[m]}(j\omega_k)\right |^2-\hat{\sigma}^2_{\hat{U},n}(j\omega_k))\\
\end{matrix}
\end{equation}
\begin{equation}\hspace{-10.59mm}
\begin{matrix}
\vspace{2mm}
\hspace{-3.073cm}\hat{S}_{Y_{0}Y_{0}}(j\omega_k)=\frac{1}{M}\sum_{m=1}^{M}\\
(\left |\hat{Y}^{[m]}(j\omega_k)\right |^2-\hat{\sigma}^2_{\hat{Y},n}(j\omega_k))\\
\end{matrix}
\end{equation}
\begin{equation}\hspace{-9.60mm}
\begin{matrix}
\hspace{-3.199cm}\hat{S}_{Y_{0}U_{0}}(j\omega_k)=\frac{1}{M}\sum_{m=1}^{M}\vspace{2mm}\\
(\hat{Y}^{[m]}\overline{\hat{U}^{[m]}}(j\omega_k)-\hat{\sigma}^2_{\hat{Y}\hat{U},n}(j\omega_k))\\
\end{matrix}
\end{equation}
\vspace{-2mm}
\begin{equation}
\begin{matrix}
\hspace{-2.0375cm}\hat{\sigma}^2_{\hat{G}_{BLA,n}}(j\omega_k)=\frac{\left |\hat{G}_{BLA}(j\omega_k)\right |^2}{M}\\
\hspace{1.2mm}\cdot (\frac{\hat{\sigma}^2_{\hat{Y},n}(j\omega_k)}{\hat{S}_{Y_{0}Y_{0}}(j\omega_k)}+\frac{\hat{\sigma}^2_{\hat{U},n}(j\omega_k)}{\hat{S}_{U_{0}U_{0}}(j\omega_k)}-2\text{Re}(\frac{\hat{\sigma}^2_{\hat{Y}\hat{U},n}(j\omega_k)}{\hat{S}_{Y_{0}U_{0}}(j\omega_k)}))\\
\end{matrix}
\end{equation}
\vspace{-2mm}
\begin{equation}\hspace{-1.80mm}
\hspace{1.024cm}\hat{\sigma}^2_{\hat{G}_s}(j\omega_k) = M(\hat{\sigma}^2_{\hat{G}_{BLA}}(j\omega_k) - \hat{\sigma}^2_{\hat{G}_{BLA,n}}(j\omega_k))
\end{equation}with $M$ is the number of experiments, $P$ is the number of periods,
$U^{[m,p]}(j\omega_k)$ and $Y^{[m,p]}(j\omega_k)$ are the Fourier
transform of one period of the input and output signals in one experiment
respectively (for a detailed discussion see Sec. 6.1 of \cite{Johan3}).}

\vspace{-5mm}

\section{Structure Detection\label{sec:Structure-Detection}}

By analyzing the BLA variations that take place, for varying DC and
STD (standard deviation of the $u(t)$) levels of the excitation signal%
, the internal structure of the system is detected. %
{} Applying Bussgang's theorem to the structures in Fig. \ref{Block_oriented},
the BLA is given below.

It is possible to select one out of 3 candidate model structures by
varying the excitation signal's characteristic values. For Gaussian
distributed excitation signal for all structures in Fig. \ref{Block_oriented},
and changing the DC and STD of the signal\textcolor{black}{{} \cite{Lieve}}
\textcolor{black}{for each mentioned structure, we have the following
results}
\begin{itemize}
\item Wiener-Hammerstein (WH) systems:
\end{itemize}
\begin{equation}
G_{BLA}(j\omega)=kG_1(j\omega)G_2(j\omega)
\end{equation}

In this structure 

\begin{equation}
G_1=\frac{N_1}{D_1},G_2=\frac{N_2}{D_2} \rightarrow G_{BLA}=k\frac{N_{1}N_{2}}{D_{1}D_{2}}
\end{equation}where, according to Bussgang's theorem, $k$ is the best linear approximation
of the static nonlinearity $q=f(p)$%
. By changing the DC and STD of the excitation signal, $k$ changes,
but poles and zeros of the $G_{BLA}$ don't change.
\begin{itemize}
\item Parallel Wiener-Hammerstein systems (example with 2 branches):
\end{itemize}
\begin{equation}
G_{BLA}=G_{1BLA}+G_{2BLA}
\end{equation}

Here the following equations hold\begin{equation}
\left\{\begin{matrix}
G_{11}=\frac{N_{11}}{D_{11}},G_{12}=\frac{N_{12}}{D_{12}} \rightarrow G_{1BLA} = k_{1}G_{11}(j\omega)G_{12}(j\omega)
\\
G_{21}=\frac{N_{21}}{D_{21}},G_{22}=\frac{N_{22}}{D_{22}} \rightarrow G_{2BLA} = k_{2}G_{21}(j\omega)G_{22}(j\omega)
\end{matrix}\right.
\end{equation}\begin{eqnarray}
G_{BLA}\!\!\!\!\!&=&\!\!\!\!k_{1}G_{11}(j\omega)G_{12}(j\omega)+k_{2}G_{21}(j\omega)G_{22}(j\omega)\\
&=&\!\!\!\!k_{1}\frac{N_{11}}{D_{11}}\frac{N_{12}}{D_{12}}+k_{2}\frac{N_{21}}{D_{21}}\frac{N_{22}}{D_{22}}\\
&=&\!\!\!\!\frac{k_{1}N_{11}N_{12}D_{21}D_{22}+k_{2}N_{21}N_{22}D_{11}D_{12}}{D_{11}D_{12}D_{21}D_{22}}
\end{eqnarray}where $k_1$ and $k_2$ are the BLAs of the static nonlinearity in
the first and second branches respectively. Here it can be seen that
the overall BLA has a pole set of the union of pole sets of each branch.
But the zeros are different from the zeros of branches. So by changing
the DC and RMS levels, zeros move but poles don't. The detailed identification
procedure can be found in \cite{Maarten1}.
\begin{itemize}
\item Nonlinear feedback:
\end{itemize}
As a first approximation%
\footnote{It should be noted, that the input of the SNL in the NL feedback case
is not Gaussian anymore. As long as the input's amplitude is small,
the input of the SNL block tends to be Gaussian. In this case the
BLA is the same as the $\epsilon$-approximation \cite{Johan1}.%
} we have that:

\begin{equation}
G_{BLA} \cong \frac{G_{1}(j\omega)}{1+G_{1}(j\omega)G_{BLA\textbf{Fb}}}
\end{equation}

In this case the relations are

\begin{equation}
\left\{\begin{matrix}
\hspace{-2.2 pt} G_{1} =\frac{N_{1}}{D_{1}} & \\
G_{21}=\frac{N_{21}}{D_{21}},&\hspace{-8 pt}G_{22}=\frac{N_{22}}{D_{22}}\rightarrow G_{BLA\textbf{Fb}} = kG_{21}(j\omega)G_{22}(j\omega)
\end{matrix}\right.
\end{equation}
\vspace{-4mm}

\vspace{-4mm}
\begin{eqnarray}
G_{BLA}& \cong & \frac{G_{1}(j\omega)}{1+G_{1}(j\omega)kG_{21}(j\omega)G_{22}(j\omega)}\\
       &=&\frac{\frac{N_{1}}{D_{1}}}{1+k\frac{N_{1}}{D_{1}}\frac{N_{21}}{D_{21}}\frac{N_{22}}{D_{22}}}\\
       &=&\frac{N_{1}D_{21}D_{22}}{D_{1}D_{21}D_{22}+kN_{1}N_{21}N_{22}}.
\end{eqnarray}%
Here $k$ is the BLA of the static nonlinearity in the feedback path.
It is seen that the overall BLA has zeros of the union of the zero
set of feedforward path and the pole set of the feedback path. By
varying the excitation signal's characteristics (e.g. DC and STD)
the overall zeros don't change, but the poles change. In the special
case where $G_{21}=G_{22}=1$ the zeros of the closed loop system
are the same as the zeros of feedforward path.%

The results of this section are collected in Table \ref{tab:Pole-zero-behavior}.
This table allows to obtain an idea of a possible internal NL structure
in an early phase (prior to the actual modeling) starting from BLAs.
E.g., if the estimated poles of the BLA are depending on DC or STD,
it can readily be determined that nonlinear feedback is present. %

\vspace{0mm}

\begin{table}
\vspace{-0.5mm}
\begin{centering}
\def\tablename{TABLE}
\caption{Pole zero behavior of considered structures. \label{tab:Pole-zero-behavior}} 
 \begin{tabular}{|c|c|c|} \cline{2-3}  \multicolumn{1}{c|}{} & Zeros & Poles\tabularnewline \hline  WH & Don't change & Don't change\tabularnewline \hline  WH. Parallel & Change & Don't Change\tabularnewline \hline  NL Feedback & Don't change & Change\tabularnewline \hline  \end{tabular} \par\end{centering}
\vspace{-5.5mm}
\end{table}

\vspace{0mm}

The reader should be aware that the mentioned pole-zero behaviors
are necessary conditions for the presence of the related structure,
in other words there are different structures other than above with
the same behavior of pole-zero. This is discussed in \cite{Johan1}.

\vspace{0mm}

\section{Optimizing, signal's characteristics\label{sec:Optimizing,-signal's-characteris}}

\textcolor{black}{In this section an experimental algorithm is introduced
to find a set of DC and STD levels of the input signal to have less
distorted BLAs. The first part of the proposed method uses the Central
Composite Design (CCD) \cite{Heikki} approach }%
\footnote{\textcolor{black}{Standard CCD method is explained through steps 1-6.}%
}\textcolor{black}{. Then on the saddle point of the quadratic model
the eigen-path gives the set of experiments, in which both the DC
and STD levels are simultaneously varied. }

\textcolor{black}{It is expected, that the behavior of a nonlinear
system is changed with varying DC and STD (amplitude) of the signal.
Also, by exciting the system with low-amplitude signals (STD), the
noise distortion is typically more significant in the BLA. This gives
either a hyper-bola, convex, or concave surface to the MSEs of BLAs,
which can be approximated by a quadratic model. This quadratic model
for MSEs is controlled by the DC and STD level. The eigen-path on
the extremum point of the surface gives the set of experimental points.
This gives a set of BLAs with less distortion levels. The algorithm
is as follows:}
\begin{enumerate}
\item Normalize the DC and STD according to their range %
$\Delta \text{DC}$ (or $\text{DC}_{max}-\text{DC}_{min}$) and $\Delta \text{STD}$
(or $\text{STD}_{max}-\text{STD}_{min}$):\begin{eqnarray}
\vspace{-10mm}
x_1\!\!=\!\!\!\!&\frac{\text{DC}-\text{DC}_C}{{\Delta \text{DC}}/_2}&\\
x_2\!\!=\!\!\!\!&\frac{\text{STD}-\text{STD}_C}{{\Delta \text{STD}}/_2}&
\end{eqnarray}where $\text{DC}_C$ and $\text{STD}_C$ are the center point of the
CCD experiment.%
{} Make the following experiment plan, where each row corresponds the
settings of an experiment:\vspace{-2mm}
\begin{equation*}
\hspace{-6mm} x_1 \ \ \ \ \ x_2
\end{equation*}\vspace{-12bp}
\begin{equation}
X_{plan} = \begin{bmatrix} +1 & -1 \\  +1 & +1 \\  -1 & -1\\  -1 & +1\\  \sqrt{2} & 0\\  -\sqrt{2} & 0\\  0 & \sqrt{2}\\  0 & -\sqrt{2}\\  0 & 0\\  \vdots  & \vdots \\  0 & 0 \end{bmatrix}
\begin{matrix}  &  & \\   &  & \\   &  & \\   &  & \\   &  & \\   &  & \\   &  & \\   &  & \\  \vspace{-1mm} \left.  \hspace{-2mm} \begin{matrix} \\  \\  \\  \end{matrix}\right\} l \textrm{\ times} \end{matrix}
\label{eq:MATRISS}
\end{equation}
\item At each experimental point for the DC and STD level of the excitation,
estimate the BLA and its MSE (Mean Square Error which is defined on
the total distortion $\hat{\sigma}^2_{\hat{G}_{BLA}}$):

\begin{equation}
\text{MSE} = \frac{1}{f_{k_N}-f_{k_1}}\int_{f_{k_1}}^{f_{k_{N_{ex}}}}\hat{\sigma}^2_{\hat{G}_{BLA}}df
\end{equation}

where, $f_{k_1}$, and $f_{k_{N_{ex}}}$ are the first and the last
excited frequencies respectively in the multisine signal.

\item Build a full quadratic model for MSEs, by using a linear least square
fit (regression with $X_{plan}$ as regressor and the MSEs as dependent
vector):

\vspace{-5mm}
\begin{equation}
\hat{\text{MSE}}=A_{20}x_1^2 + A_{11}x_1x_2 + A_{02}x_2^2 + A_{10}x_1 + A_{01}x_2 + A_{00}
\label{eq:MSE}
\end{equation}

\item By doing multiple ($l$) experiments at the center point, it gives
the ability to measure the noise level in MSEs. Based on that, the
uncertainty of the estimated parameters also can be estimated:

\vspace{-3mm}
\begin{equation}\hspace{-16mm}
\text{RSS} = \sum_{i=1}^{n}({\text{MSE}}_{m_{i}}- \hat{\text{MSE}_i} )
\end{equation}
\vspace{-3mm}
\begin{equation}\hspace{-37.19mm}
\text{Var}(\text{MSE})=\text{RSS}/(n-p)
\end{equation}
\vspace{-5mm}
\begin{equation}\hspace{-21.16mm}
\text{Cov}( \hat{A}_i )=(X_m^TX_m)^{-1}\text{Var}(\text{MSE})
\end{equation}where \text{RSS} is the residual sum of squares, $\hat{\text{MSE}}$
is the vector of predicted $\text{MSE}$s, the subscript $m$ represents
the measured values of each variable, $n$ is the number of all experiments,
and $p$ is the degree of freedom (number of unknown parameters $A_i$),
that here is equal to the number of parameters in the model.

\item Calculate the signal to noise ratio ($t_i$) for each coefficient
according to:

\vspace{-2mm}
\begin{equation}
t_i=\frac{\hat{A}_i}{\text{STD}{\hat{A}_i}}
\end{equation}

where $\hat{A}_i$ is the estimated coefficients and $\text{STD}{\hat{A}_i}$
is:

\begin{equation}
\text{STD} ( \hat{A}_i )=\sqrt{diag ( \text{Cov} ( \hat{A}_i ) ) }
\end{equation}

\item Drop the coefficients with low absolute signal to noise ratio ($\left|t_i\right|$).
As a rule of thumb, it is recommended to drop the coefficients with
$t$-values less than 3,
\item Find the extremum point of the surface by solving:

\begin{equation}
\begin{bmatrix} 2A_{20} & A_{11}\\  A_{11} & 2A_{02} \end{bmatrix} \begin{bmatrix} x^*_{1}\\  x^*_{2} \end{bmatrix} + \begin{bmatrix} A_{10}\\  A_{01} \end{bmatrix} = \begin{bmatrix} 0\\  0 \end{bmatrix}
\end{equation}

\item Transfer (\ref{eq:MSE}) to the new point $(x^*_{1},x^*_{2})$
\item This new equation can be written as:

\vspace{-5mm}
\begin{equation*}
\hspace{20mm}Q
\vspace{-4mm}
\end{equation*}
\begin{eqnarray}
\text{MSE}_{ex} \hspace{-3mm} &=& \hspace{-3mm} 
\begin{bmatrix} \tilde{x}_{1} & \tilde{x}_{2} \\ \end{bmatrix} \overbrace{\begin{bmatrix} \tilde{A}_{20} & \tilde{A}_{11}/_2 \\ \tilde{A}_{11}/_2 & \tilde{A}_{02} \\ \end{bmatrix}} \begin{bmatrix} \tilde{x}_{1}\\\tilde{x}_{2}\\ \end{bmatrix}\\
\hspace{-3mm} &=& \hspace{-3mm}  \tilde{A}_{20}\tilde{x}_{1}^2 + \tilde{A}_{11}\tilde{x}_{1}\tilde{x}_{2} + \tilde{A}_{02}\tilde{x}_{2}^2
\end{eqnarray}

where $\tilde{x}_{i}=x_i-x^*_i(i=1,2)$ is a new transferred variable.

\item The eigenvector of the matrix $Q$ corresponding to the lowest eigenvalue
determines the line (through $x^*$) along which the MSE varies the
least, and therefore delivers different BLAs of the highest quality.
This gives the best strategy for doing the experiments.
\end{enumerate}
\vspace{-3mm}

\section{Systems under study\label{sec:Systems-under-study}}

\begin{figure}[tbh]
\begin{centering}
\includegraphics[clip=0bp 0bp 412bp 390bp,scale=0.3]{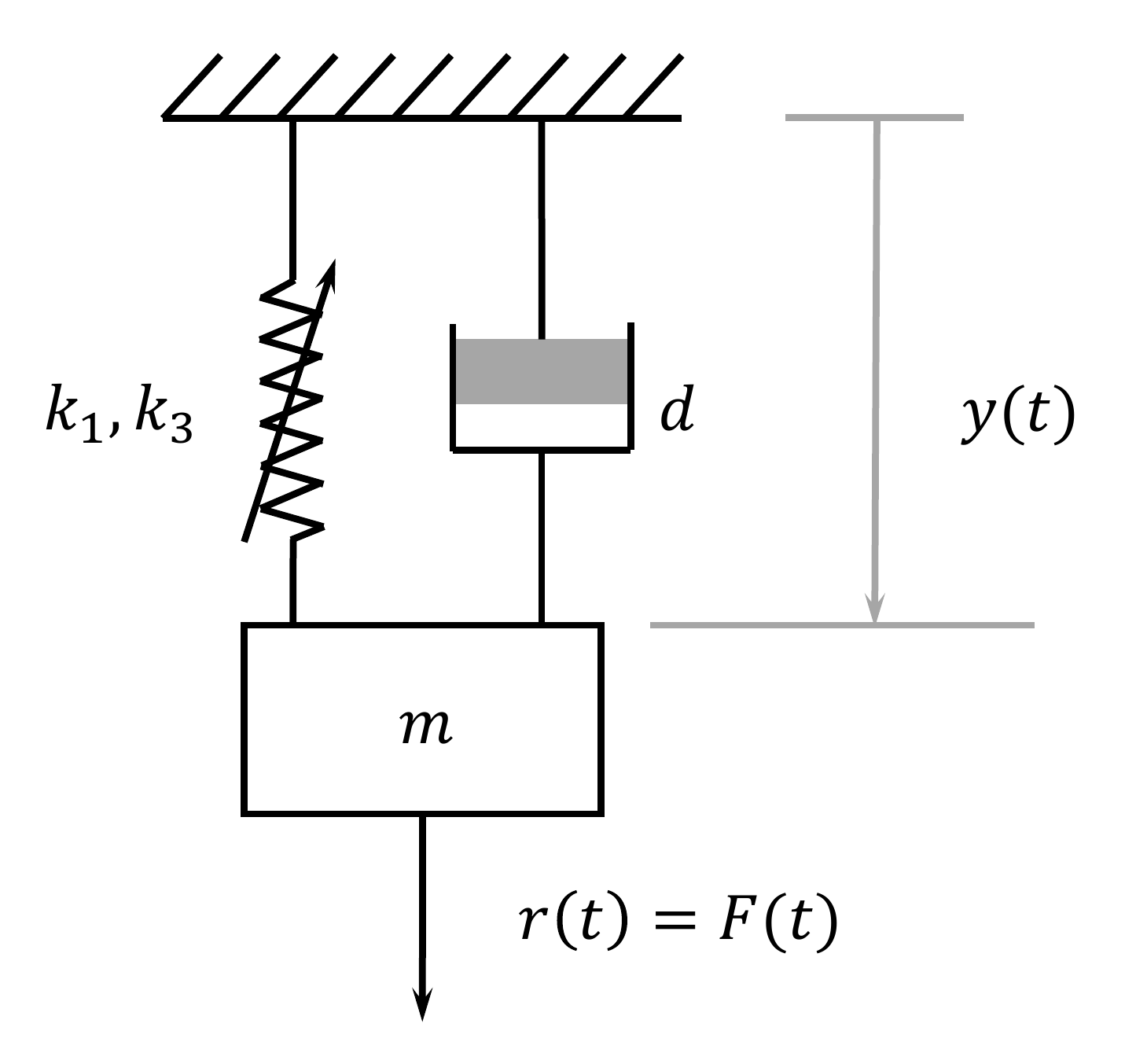}
\par\end{centering}

\caption{Nonlinear mass spring damper with a cubic term as hardening effect.\label{Nonlinear-mass-spring}}
\end{figure}

Two nonlinear feedback systems are considered in this study, a nonlinear
mass spring damper (NL-MSD) system (see Fig. \ref{Nonlinear-mass-spring}
and Fig. \ref{Block-diagram-of}$(a)$), and a linear system with
a multiplication operation in the feedback path (NL-XFB) (Fig. \ref{Block-diagram-of}$(b)$).
Both systems are implemented as electronic circuits. The proposed
methodology will be applied to measurements of these devices (see
Section \ref{sec:Results}).

\subsection{The nonlinear mass spring damper system (NL-MSD)}

This system has a spring with a nonlinear cubic term $k_{3}y^3$ to
model a mechanical hardening spring effect. For this system we have
\begin{eqnarray}
m\ddot{y}+d\dot{y}+k_{1}y+k_{3}y^3&=&r(t)\\
m\ddot{y}+d\dot{y}+k_{1}y&=&r(t)-k_{3}y^3
\end{eqnarray}

The equivalent block diagram of this system is shown in Fig. \ref{Block-diagram-of}$(a)$
(also known as the Silverbox) \cite{Johan6}.

\subsection{The nonlinear system with a multiplication in the feedback path (NL-XFB)}

\begin{figure}[tbh]
\includegraphics[scale=0.29]{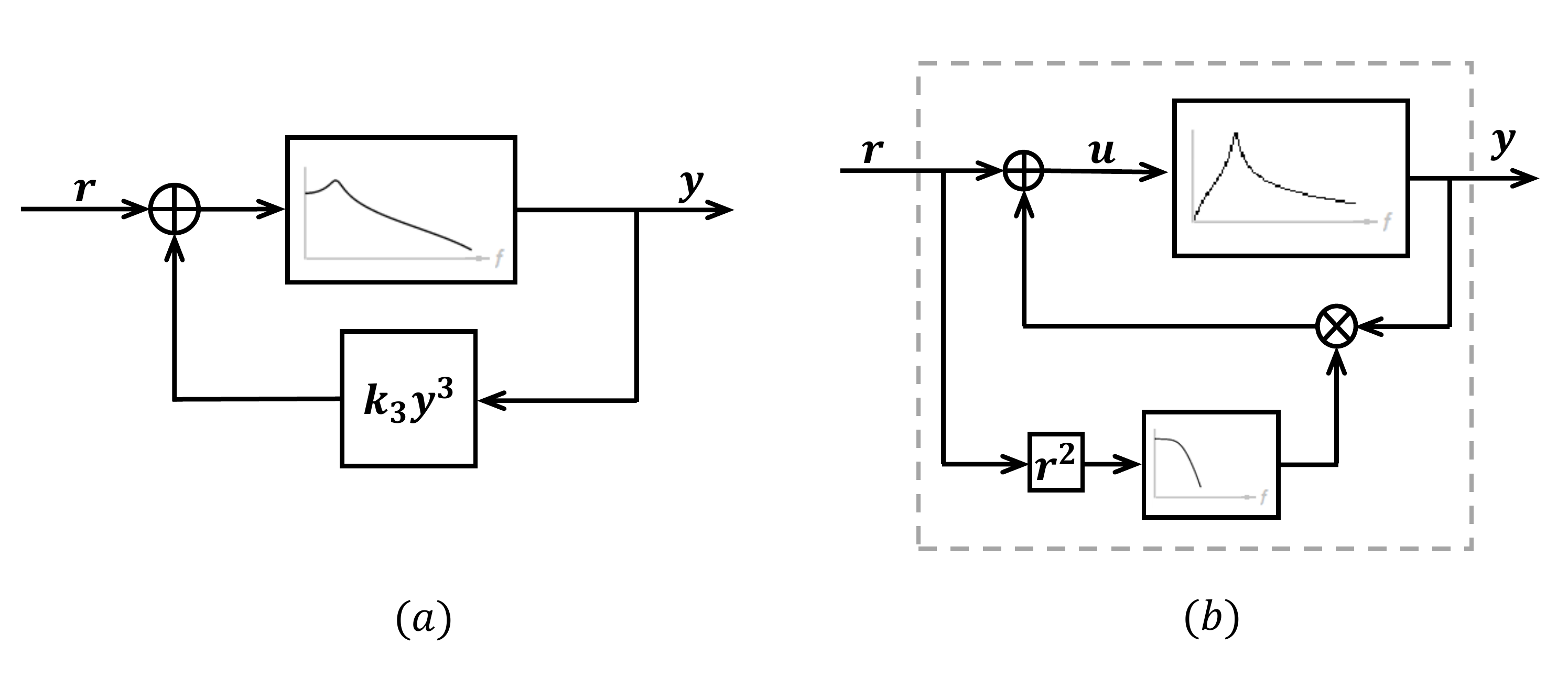}
\caption{Block diagram of systems under study. \label{Block-diagram-of}}
\end{figure}

The second system, shown in Fig. \ref{Block-diagram-of}$(b)$ is
composed of a dual amplifier bandpass filter \cite{Zumba} in the
forward path with a multiplication at the feedback which is fed by
an independent branch from the input with a squaring device (AD835) followed by a generalized impedance
converter (GIC)-derived dual-amplifier biquad as a low-pass filter
\cite{Chen}.

\section{Measurement Results\label{sec:Results}}

\textcolor{black}{Two systems under study are excited with random
phase multisine signal. The settings for the excitation signal are
written in Table \ref{Experiment-parameters}, and measured with Agilent/HP
E1430A data acquisition cards. Both systems are excited by an Agilent/HP
E1445A arbitrary wave generator (AWG).}

\subsection{Structure Detection}

\begin{centering}
\begin{figure}
\textsf{\includegraphics[trim=20mm 40mm 0 0,scale=0.37]{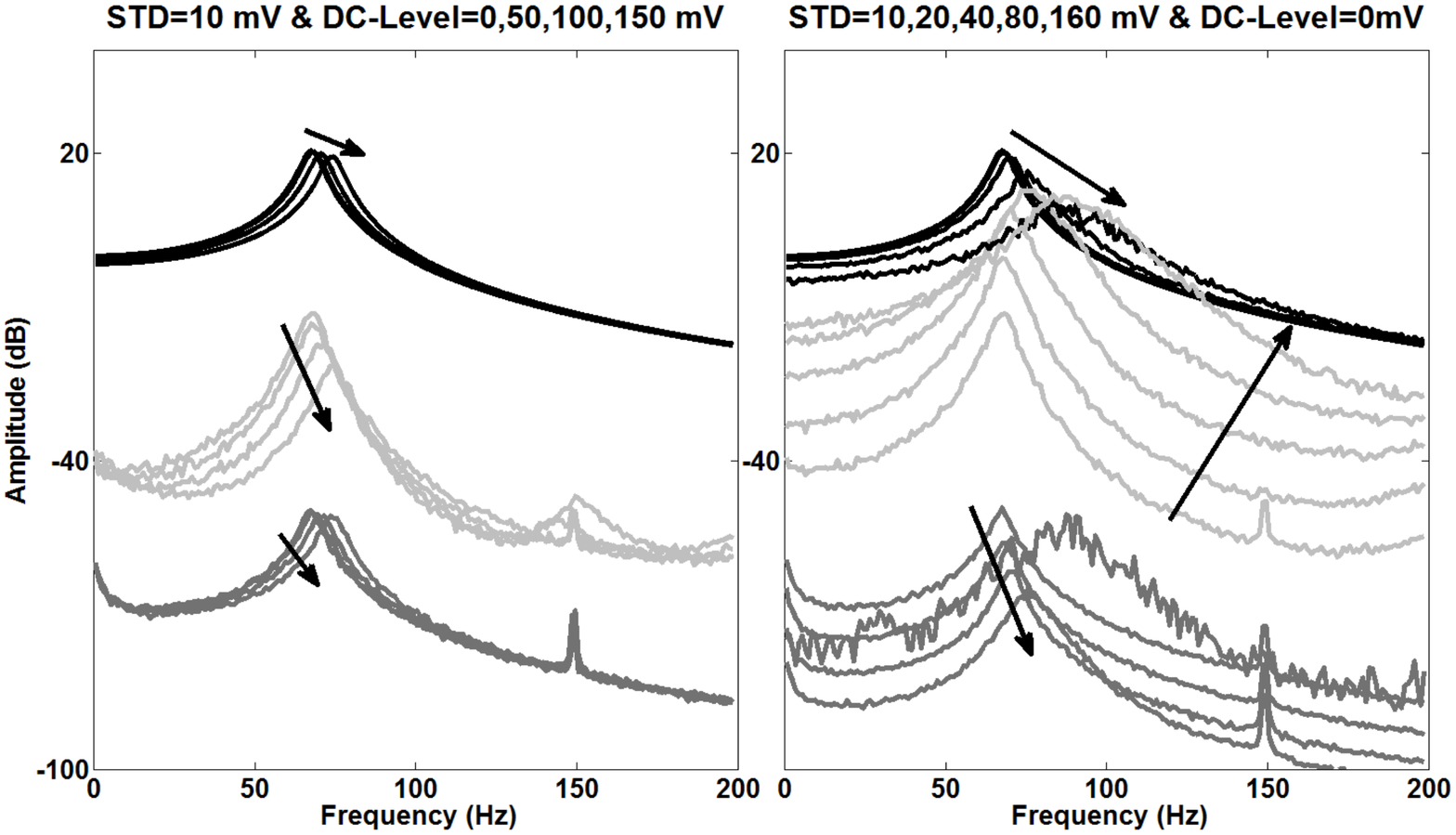}}
\caption{FRFs of the NL-MSD (Silverbox). Bold line: $\hat{G}_{BLA}$, light
grey: $\hat{\sigma}_{\hat{G}_{BLA}}$, and dark grey: $\hat{\sigma}_{\hat{G}_{BLA,n}}$.
Arrows point from low to high DC or STD levels.\label{SilverBox_FRFs}}
\end{figure}
\end{centering}

The FRFs (BLAs) of the NL-MSD are plotted in Fig. \ref{SilverBox_FRFs}.
In the left of this figure, FRFs with different DC level of the excitation
signals are shown, while in the right, FRFs are plotted for different
STD levels. It is seen that by increasing the input amplitude (STD),
the resonance frequency shifts and also the damping changes. This
behavior points to pole movements, hence to the presence of a nonlinear
feedback \cite{Johan5}.

It is seen (in the left) that the total standard deviation ($\hat{\sigma}_{\hat{G}_{BLA}}$)
and the noise level ($\hat{\sigma}_{\hat{G}_{BLA,n}}$) are decreased,
by increasing the DC level of the input signal. As the DC level is
increased, the total standard deviation and noise level are decreased,
but as the STD level is increased (see right side of the Fig. \ref{SilverBox_FRFs}),
the stochastic nonlinear distortion ($\hat{\sigma}_{\hat{G}_{BLA,n}}$)
is increased, (as expected, high amplitude hits the nonlinearities
more).

\begin{figure}
\includegraphics[trim=20mm 40mm 0 0 ,scale=0.37]{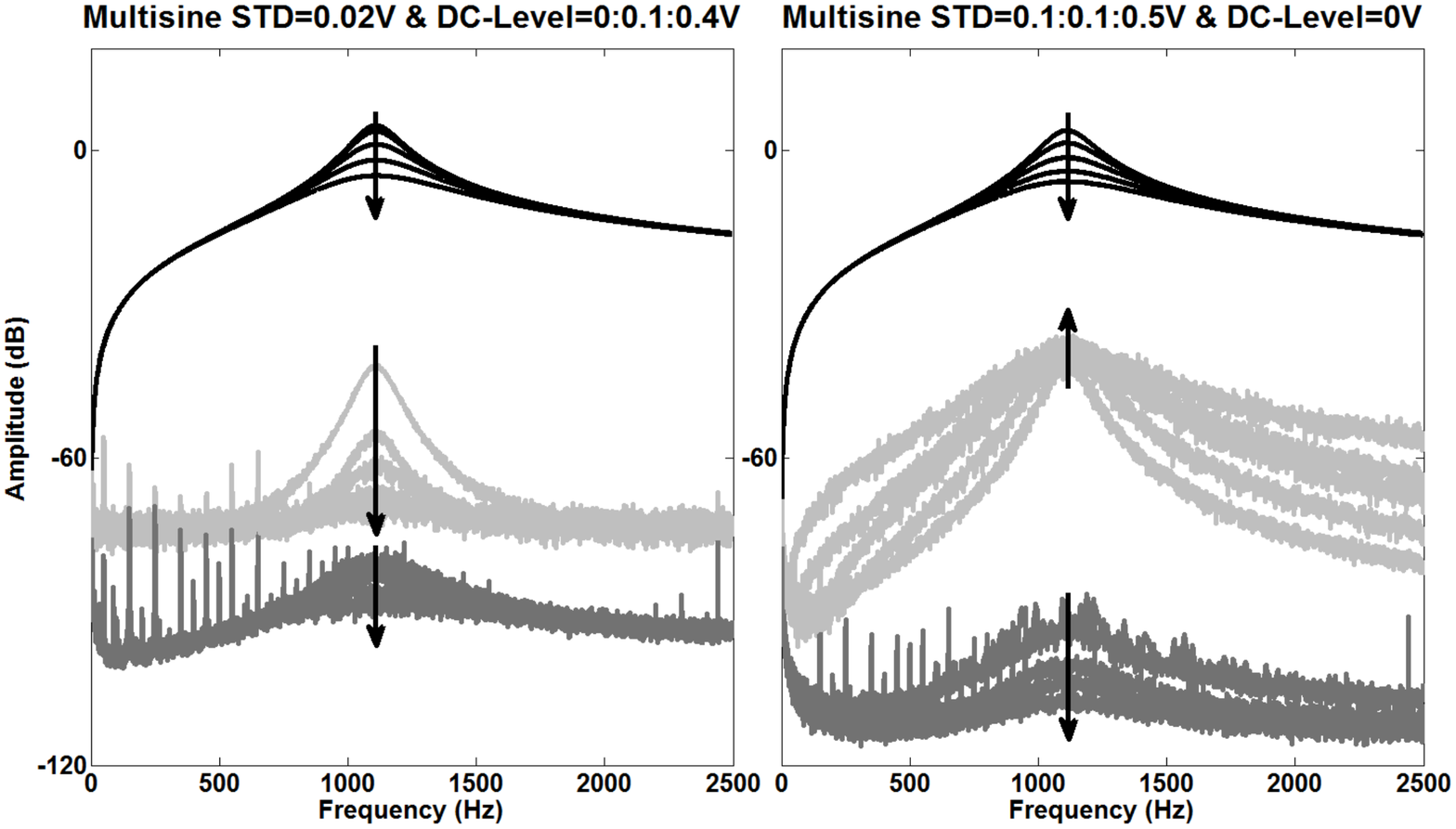}
\caption{FRFs of the NL-XFB. Arrows point from low to high DC or STD levels.
Bold line: $\hat{G}_{BLA}$, light grey: $\hat{\sigma}_{\hat{G}_{BLA}}$,
and dark grey: $\hat{\sigma}_{\hat{G}_{BLA,n}}$. \label{Whitebox_FRFs}}
\end{figure}

Fig. \ref{Whitebox_FRFs} shows a similar picture for the NL-XFB system.
In the left side of this figure there are plots of the FRFs with varying
the DC level. At the right side, the FRFs are calculated with the
different STD levels of the excitation signal. In the varying DC level
experiment, the total distortion is decreased, while in the STD sweeping
experiment the total distortion has an increasing trend. Therefore
it is recommended to use rather the DC sweeping than the varying STD,
to detect the internal structure. 

\begin{table}[tbh] \begin{centering} 
\def\tablename{TABLE}
\caption{Experiment parameters. \label{Experiment-parameters}} 
\begin{tabular}{|c|c|c|} \cline{2-3}  \multicolumn{1}{c|}{} & NL-MSD & NL-XFB\tabularnewline \hline  Number of points (per period) $(N)$ & 4883 & 19531\tabularnewline \hline  Number of realizations $(M)$ & 64 & 16\tabularnewline \hline  Number of periods $(P)$ & 4 & 4\tabularnewline \hline  Sampling frequency $(f_{s})$ & 2.44 KHz & 9.77 KHz\tabularnewline \hline  Excited harmonics $k_{ex}$ & \text{[3:2:399]} & \text{[1:2:4999]}\tabularnewline \hline  
Number of excited frequencies ($N_{ex}$) & \text{199} & \text{2500} \tabularnewline \hline
frequency resolution ($f_0$) & \text{0.5Hz} & \text{0.5Hz} \tabularnewline \hline
\end{tabular} \par\end{centering}
\end{table}

\begin{figure}
\includegraphics[trim=25mm 40mm 0 0,scale=0.36]{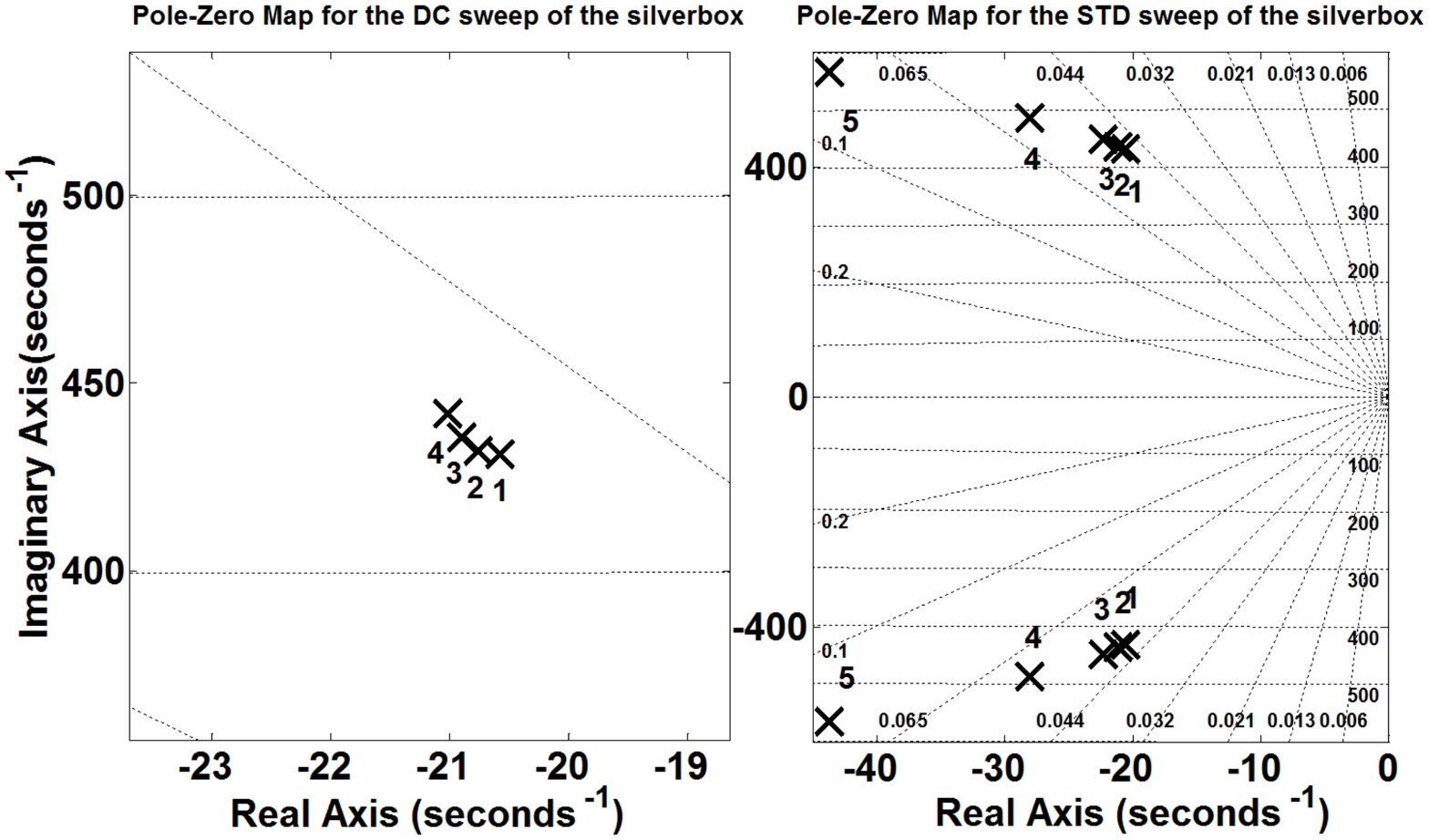}

\caption{Pole-Zero map of varying DC (left) and varying STD experiments on
the NL-MSD system. Labels 1 to 4 correspond to DC and STD levels shown
in Fig. \ref{SilverBox_FRFs} (low to high)\label{PZ_map_SilB}.}
\end{figure}

\begin{figure}
\includegraphics[trim=25mm 40mm 0 0,scale=0.36]{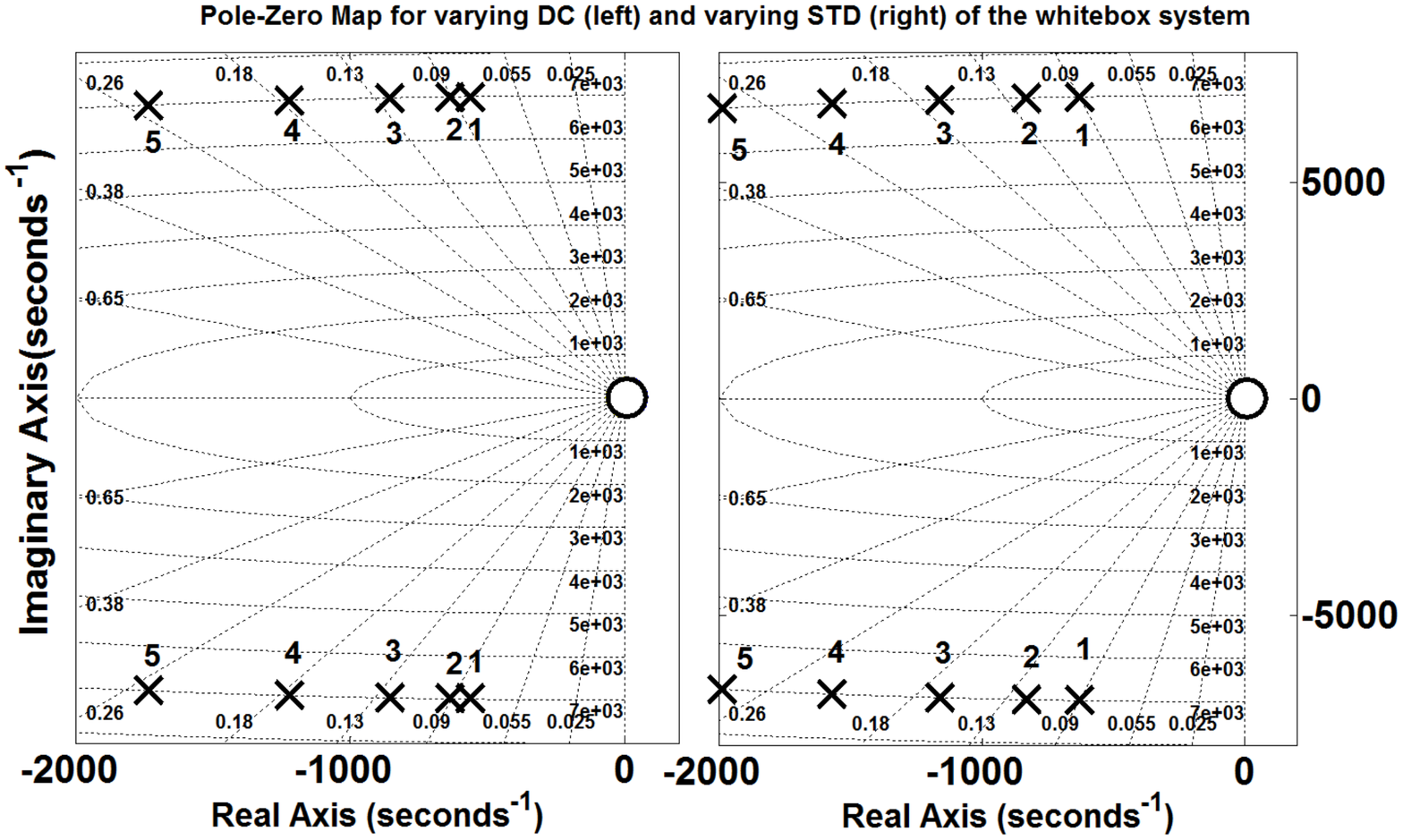}

\caption{Pole-Zero map of varying DC (left) and varying STD experiments on
the NL-XFB system. Labels 1 to 5 correspond to DC and STD levels shown
in Fig. \ref{Whitebox_FRFs} (low to high)\label{PZ_map_WB}.}
\end{figure}

Through parametric fitting on frequency response functions, by using
$\text{Matlab}^\circledR$ FDIDENT toolbox \cite{Istvan}, the movement
of poles and zeros can be followed. It is assumed that the NL-MSD
system has second order dynamics (without zeros), and the NL-XFB is
a second order system with a zero (see the deep antiresonance in Fig.
\ref{Whitebox_FRFs} at the frequency 0Hz). In Figs. \ref{PZ_map_SilB}
and \ref{PZ_map_WB} the pole-zero locus of systems under study are
illustrated.
In Fig. \ref{PZ_map_WB} the movement of poles have all the same
natural frequency. As it is seen in Figs. \ref{PZ_map_SilB} and \ref{PZ_map_WB}
the poles move in both systems. So, the presence of the nonlinear
feedback can be detected in both systems.

In Fig. \ref{PZ_map_SilB} the pole-zero map of the NL-MSD system
by a DC sweep (left), and a STD sweep (right) show pole movements
of the system. This behavior suggests again the existence of the feedback
branch. Also in Fig. \ref{PZ_map_WB}, in both cases (STD and DC sweep),
the poles are shifting, while the zeros don't change. This behavior
allows again one to easily detect the existence of the feedback branch.

\vspace{-5mm}

\subsection{Input Optimizaion}

In this part the results of the proposed method for selecting the
DC and STD level of the signal are shown. These results are based
on the experiments carried out on the NL-MSD system. 

\subsubsection{Experiments on a grid in the DC-STD plane}

The 11 DC and 16 STD levels are selected according to Table \ref{tab:DC_STD}\textcolor{black}{.
For this set of experiments the sampling frequency, number of excited
frequency, and the frequency resolution are selected according to
the NL-MSD column of Table \ref{Experiment-parameters}. Fig. \ref{fig:MSE_ALL_dB}
shows the MSEs (in dB) of all signals with permissible DC and STD
levels of the excitation signal. The results agree with the fact that,
by decreasing the STD level, the MSE is decreased because, the noise
has more contribution in the total distortion. But by increasing the
STD level, the MSEs (dB) are increased, because of the nonlinear contribution
of the system. It is clear that varying the DC level of the input
signal at a fixed STD level, doesn't have a significant impact on
MSE levels.}

\subsubsection{\textcolor{black}{Proposed method}}

\begin{figure}[tbh]
\includegraphics[trim=70mm 65mm 0 0,clip ,scale=0.55]{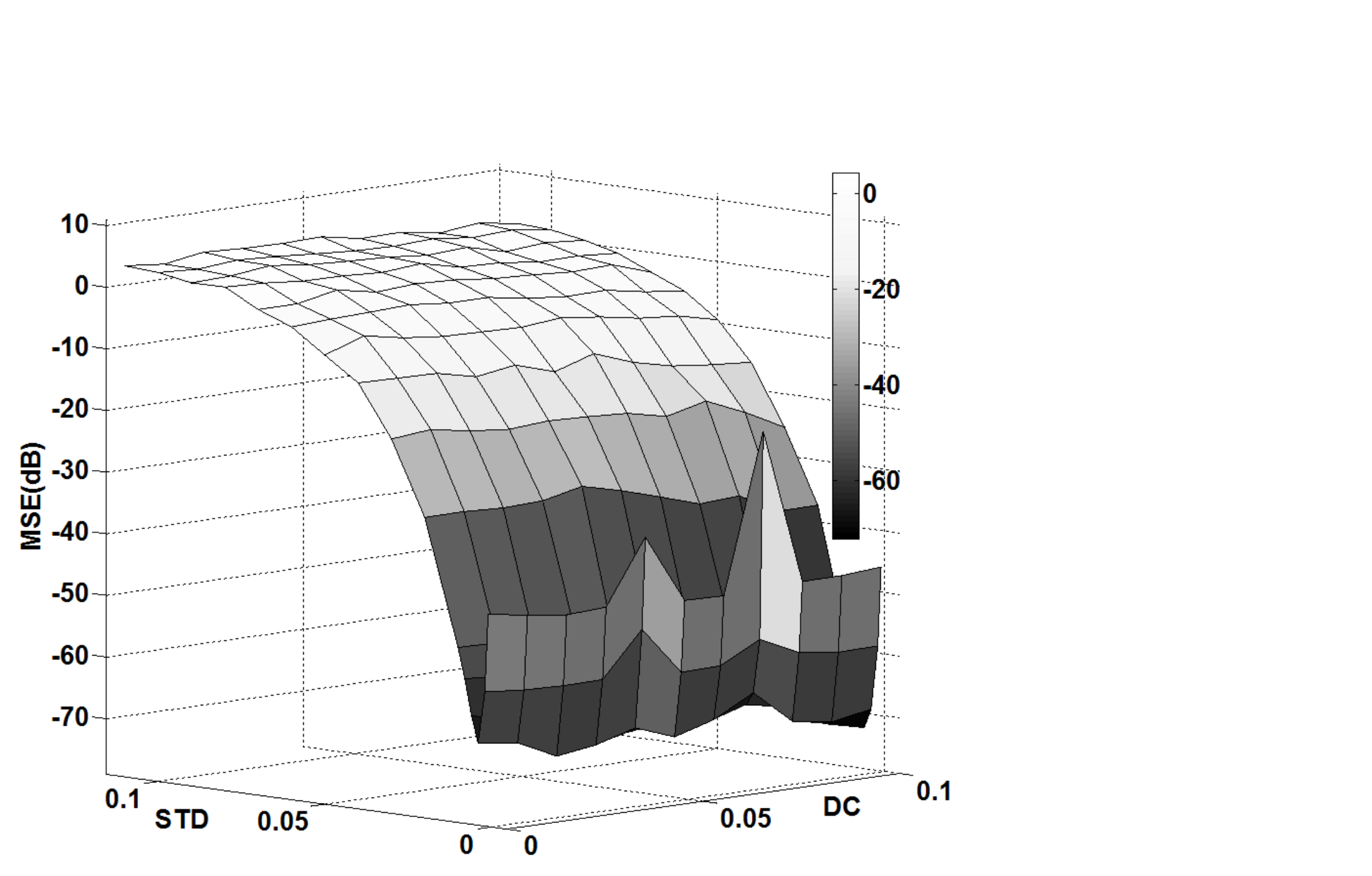}
\caption{Surface $\text{MSE} ( \text{DC},\text{STD} )$ for the NL-MSD system
(DC and STD levels according to Table \ref{tab:DC_STD}): MSE in dB
as function of the DC value and the STD. \label{fig:MSE_ALL_dB}}
\end{figure}

\begin{figure}[!t]
\graphicspath{{Figures/}}
\includegraphics[trim=23mm 40mm 0 0,scale=0.367]{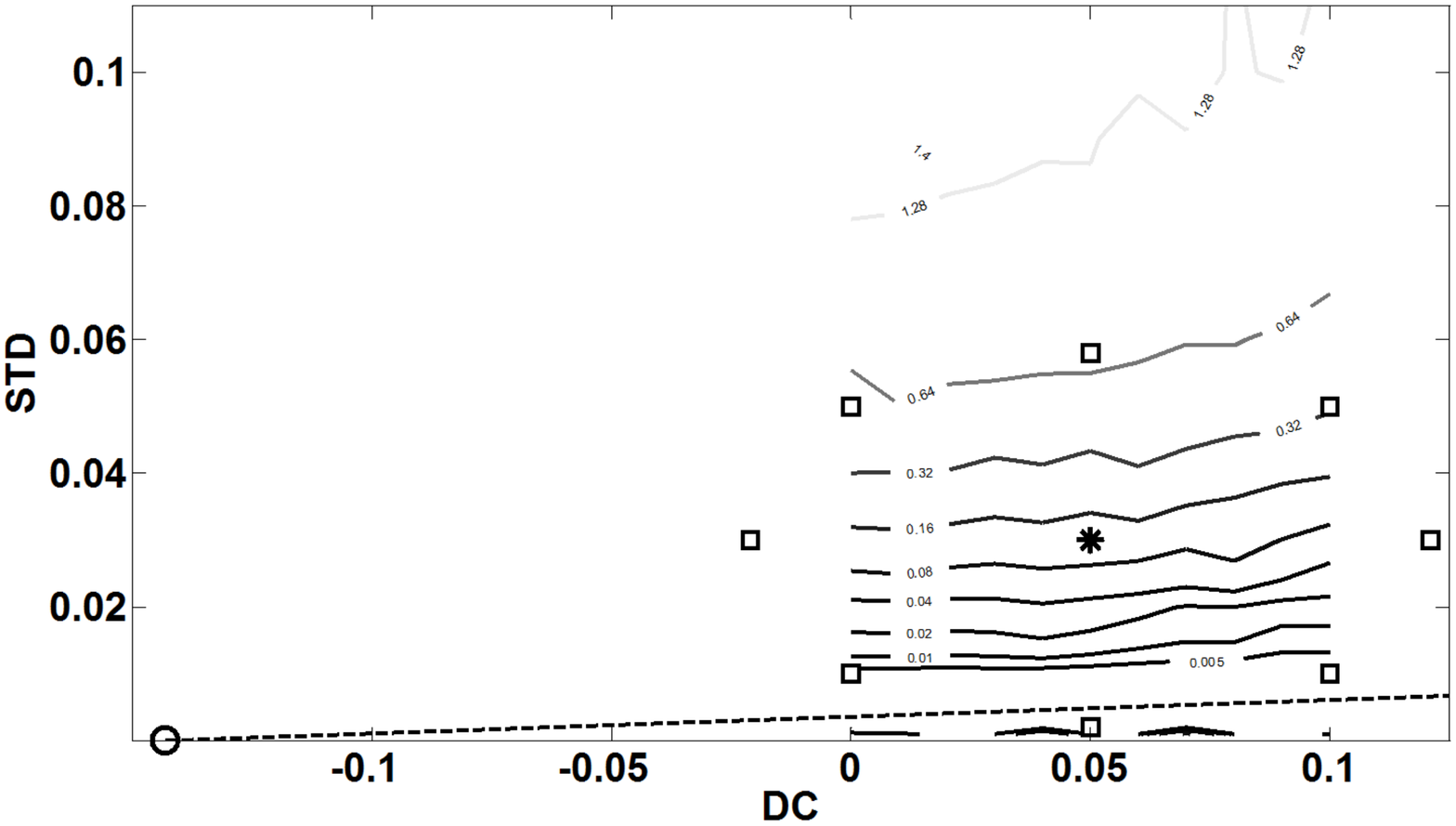}
\caption{Contour plot of MSEs for grid experiments (Fig. \ref{fig:MSE_ALL_dB}).
Here MSE scale is linear. Squares ($ \square $) and the asterisk
($ \ast $) are the CCD designed experiment points. The circle ($\circ$)
at the far left bottom of the figure is the calculated extremum point
of the quadratic model. The dash-line is the eigen-path which is selected
as the best strategy.\label{fig:MSE_CONT}}
\end{figure}

\begin{figure}
\includegraphics[trim=10mm 40mm 0 0,scale=0.45]{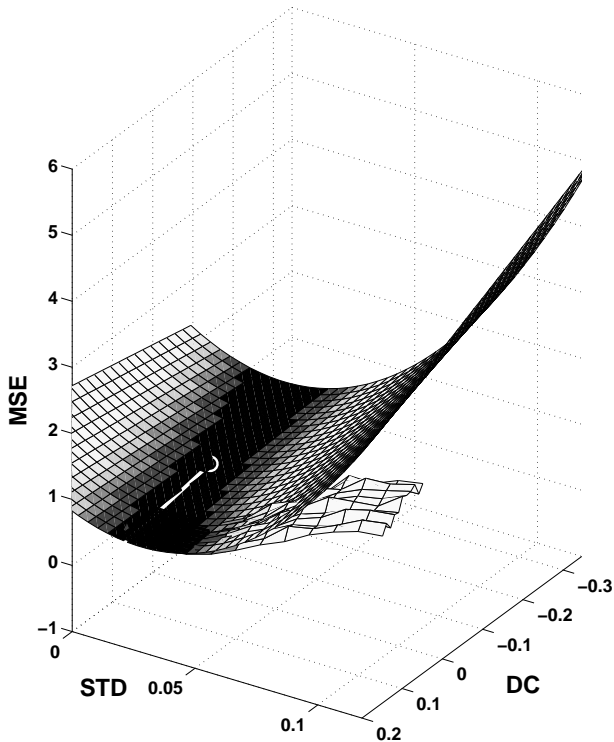}

\caption{The extracted $ 2^{\text{nd}} $ order surface for the MSEs (with
fine meshes), the grid experiment surface (with coarse meshes), the
eigen-path (the white line) and the saddle point of the modeled surface
(the white circle). \label{fig:CCD_SURF}}
\end{figure}

The CCD method explained in Section \ref{sec:Optimizing,-signal's-characteris}
was applied to find the best strategy to reach to the minimum value
of MSEs, based on only 10 different values of (DC, STD). Fig. \ref{fig:MSE_CONT}
shows the contour plot of all the experiments grid (Fig. \ref{fig:MSE_ALL_dB}).
In this figure squares and the star show the designed experiments
according to the matrix in (\ref{eq:MATRISS}). The quadratic surface
which is built according to these experiments is plotted in Fig. \ref{fig:CCD_SURF}.
The dash-line in Fig. \ref{fig:MSE_CONT} shows the best strategy
we should select to reach to an FRF with a minimum distortion level.
Also this line is shown in Fig. \ref{fig:CCD_SURF}, as a white line.

\begin{figure}[tbh]
\includegraphics[clip=24bp 0bp 900bp 612bp,scale=0.35]{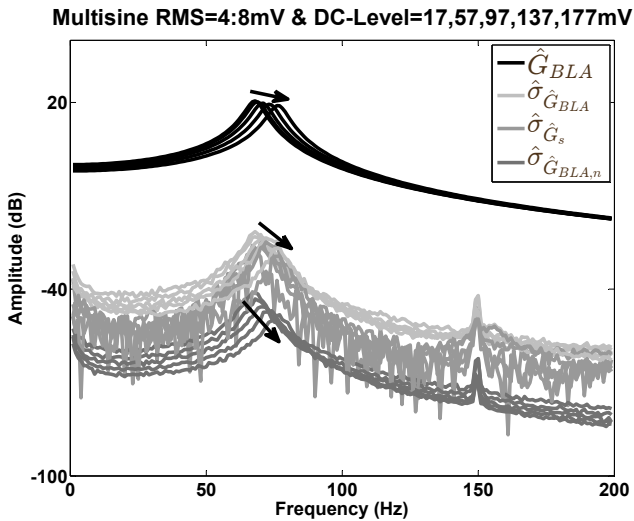}

\vspace{-8mm}

\caption{FRFs of the designed excitation signal. Arrows point from the lowest
value of DC and STD levels to the highest. Bold line: $\hat{G}_{BLA}$,
light grey: $\hat{\sigma}_{\hat{G}_{BLA}}$, dark grey: $\hat{\sigma}_{\hat{G}_{BLA,n}}$,
and medium grey is the stochastic nonlinear distortion $\hat{\sigma}_{\hat{G}_{s}}$.
\label{fig:Optimal-FRFs-of-the}}
\end{figure}

\begin{figure}
\begin{centering}
\includegraphics[trim=20mm 70mm 0 0 ,scale=0.363]{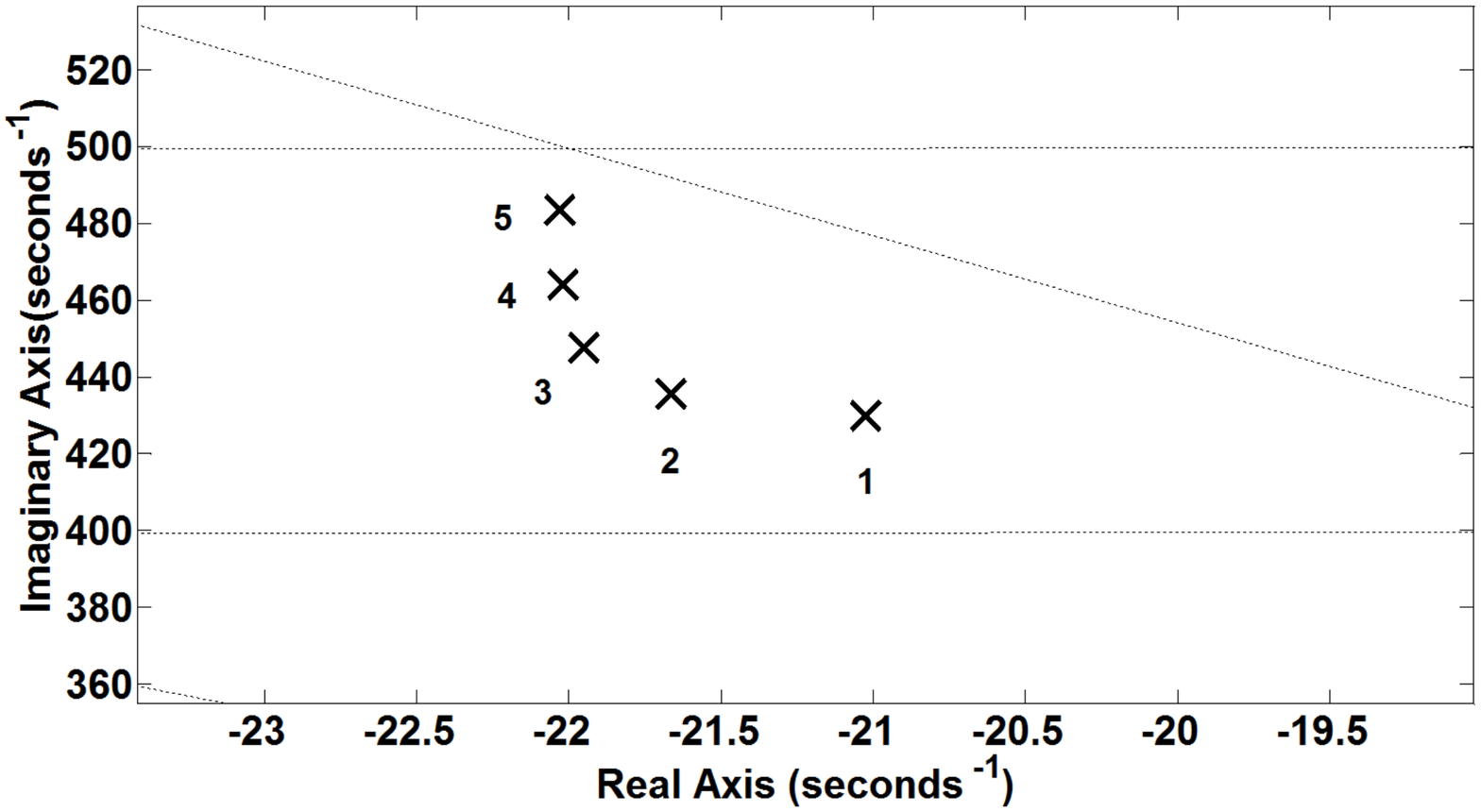}
\end{centering}
\caption{Pole-Zero map of the designed experiment for the NL-MSD system.\label{fig:P-Z-optim}}
\end{figure}

The designed DC and STD levels according to this line are in Table
\ref{tab:The-designed-experiment}. Fig. \ref{fig:Optimal-FRFs-of-the}
shows the FRFs of the designed experiment. By comparing this figure
with the Fig. \ref{SilverBox_FRFs} (the left side, FRFs for varying
DC), the noise contribution is slightly increased but the total and
stochastic nonlinear distortions decrease significantly. Moreover,
the maximum of the total distortion only varies very slightly, corresponding
to slight changes of the MSE on the optimal line. By fitting a parametric
model on each FRFs in the optimal experiment we have the poles-zeros
of transfer functions. Fig. \ref{fig:P-Z-optim} shows the pole-zero
trend of the new experiment.

\begin{table}[tbh]

\def\tablename{TABLE}

\caption{Settings for full grid measurement}\label{tab:DC_STD}
\begin{tabular}{|c|c|}
\hline 
DC Levels & 0, 10, 20, 30, 40, 50, 60, 70, 80, 90, 100 mV\tabularnewline
\hline 
\multirow{2}{*}{STD Levels} & 1, 2, 4, 6, 8, 10, 20, 30, 40, 50, 60, 70, 80, 90, 100,\tabularnewline
 &  110 mV\tabularnewline
\hline 
\end{tabular}
\end{table}

\begin{table}[tbh]
\caption{The designed experiment values (results of CCD method)\label{tab:The-designed-experiment}}
\centering{}%
\begin{tabular}{|c|c|}
\hline 
DC Levels & 17, 57, 97, 137, 177 mV\tabularnewline
\hline 
STD Levels & 4, 5, 6, 7, 8 mV\tabularnewline
\hline 
\end{tabular}
\end{table}

\section{Conclusions \label{sec:Conclusions}}

Bussgang's theorem is still valid for nonzero mean excitation signals.
By using the Input/Output BLA of a system, with varying input signal
properties (DC and STD level), and studying the pole zero behavior
of the BLA, we are able to %
select a block oriented structure out of the 3 different types of
internal structures i.e. series, parallel Wiener-Hammerstein systems,
and nonlinear feedback structure (see Table \ref{tab:Pole-zero-behavior}).
By changing the DC and STD levels, it is possible to influence the
level of the nonlinear and noise distortions. Varying the DC level
is preferable to varying the STD, since delivers much higher quality
(lower distortion) BLA measurements. An optimal experimental design
procedure is proposed, where DC and STD values are varied jointly.

\appendices{%
}

\bibliographystyle{IEEEtran}
\bibliography{BIBLIO2}

\begin{thebibliography}{10}
\providecommand{\url}[1]{#1}
\csname url@samestyle\endcsname
\providecommand{\newblock}{\relax}
\providecommand{\bibinfo}[2]{#2}
\providecommand{\BIBentrySTDinterwordspacing}{\spaceskip=0pt\relax}
\providecommand{\BIBentryALTinterwordstretchfactor}{4}
\providecommand{\BIBentryALTinterwordspacing}{\spaceskip=\fontdimen2\font plus
\BIBentryALTinterwordstretchfactor\fontdimen3\font minus
  \fontdimen4\font\relax}
\providecommand{\BIBforeignlanguage}[2]{{%
\expandafter\ifx\csname l@#1\endcsname\relax
\typeout{** WARNING: IEEEtran.bst: No hyphenation pattern has been}%
\typeout{** loaded for the language `#1'. Using the pattern for}%
\typeout{** the default language instead.}%
\else
\language=\csname l@#1\endcsname
\fi
#2}}
\providecommand{\BIBdecl}{\relax}
\BIBdecl

\bibitem{Billings}
S.~A. Billings and S.~Y. Fakhouri, ``Identification of systems containing
  linear dynamic and static nonlinear elements,'' \emph{Automatica}, vol.~18,
  pp. 15--26, 1982.

\bibitem{Fouad}
F.~Giri and E.-W. Bai, \emph{Block-oriented Nonlinear System
  Identification}.\hskip 1em plus 0.5em minus 0.4em\relax Springer-Verlag,
  2010.

\bibitem{Billings3}
S.~A. Billings and K.~M. Tsang, ``Spectral analysis of block structured
  non-linear systems,'' \emph{Mechanical Systems and Signal Processing},
  vol.~4, pp. 117--130, 1990.

\bibitem{Haber1}
R.~Haber and H.~Unbehauen, ``Structure identification of nonlinear dynamic
  systems-a survey on input/output approaches,'' \emph{Automatica}, vol.~26,
  pp. 651--677, 1990.

\bibitem{Rik1}
R.~Pintelon and J.~Schoukens, \emph{System Identification a Frequency Domain
  Approach}.\hskip 1em plus 0.5em minus 0.4em\relax John Wiley \& Sons, Inc.,
  2012.

\bibitem{Billings2}
S.~A. Billings and S.~Y. Fakhouri, ``Identification of non-linear unity
  feedback systems,'' University of Sheffield, Tech. Rep., 1978.

\bibitem{Lieve}
L.~Lauwers, J.~Schoukens, R.~Pintelon, and M.~Enqvist, ``A nonlinear block
  structure identification procedure using frequency response function
  measurements,'' \emph{IEEE Trans. Instrum. Meas.}, vol.~57, pp. 2257--2264,
  2008.

\bibitem{khodam}
F.~E. Alireza, J.~Schoukens, and L.~Vanbeylen, ``Design of excitations for
  structure discrimination of nonlinear systems, using the best linear
  approximation,'' in \emph{Proc. IEEE Instrum. Meas., Technol. Conf.}, May
  2015, pp. 234--239.

\bibitem{Heikki}
W.~G.~H. George E. P.~Box, J. Stuart~Hunter, \emph{Statistics for
  experimenters: design, discovery, and innovation}.\hskip 1em plus 0.5em minus
  0.4em\relax John Wiley \& Sons, Inc., 2005.

\bibitem{Bussgang}
J.~J. Bussgang, ``Crosscorrelation functions of amplitude-distorted gaussian
  signals,'' Res. Lab. Electron., Mass. Inst. Technol.,Cambridge, MA, MIT Tech
  Rep. P. No 216, Tech. Rep., 1952.

\bibitem{Johan1}
J.~Schoukens, R.~Pintelon, Y.~Rolain, M.~Schoukens, K.~Tiels, L.~Vanbeylen,
  A.~V. Mulders, and G.~Vandersteen, ``Structure discrimination in
  block-oriented models using linear approximations: A theoretic framework,''
  \emph{Automatica}, vol.~53, pp. 225--234, 2015.

\bibitem{Nuttall}
A.~H. Nuttall, ``Theory and application of the separable class of random
  processes,'' MIT, Tech. Rep., 1958.

\bibitem{Johan2}
J.~Schoukens, L.~Gomme, W.~V. Moer, and Y.~Rolain, ``Identification of
  block-structured nonlinear feedback system, applied to a microwave crystal
  detector,'' \emph{IEEE Trans. Instrum. Meas.}, vol.~57, pp. 1734--1740, 2008.

\bibitem{Bendat}
J.~S. Bendat and A.~G. Piersol, \emph{Engineering Application of Correlation
  and Spectral Analysis}.\hskip 1em plus 0.5em minus 0.4em\relax New York, NY,
  USA,:Wiley, 1980.

\bibitem{Johan4}
J.~Schoukens, R.~Pintelon, T.~Dobrowiecki, and Y.~Rolain, ``Identification of
  linear systems with non-linear distortions,'' \emph{Automatica}, vol.~41, pp.
  491--504, 2005.

\bibitem{Enqvist2}
M.~Enqvist and L.~Ljung, ``Linear approximations of nonlinear fir systems for
  separable input processes,'' \emph{Automatica}, vol.~41, pp. 459--473, 2005.

\bibitem{Rik2}
R.~Pintelon and J.~Schoukens, ``Measurement and modelling of linear systems in
  the presence of non-linear distortions.'' \emph{Mechanical systems and signal
  processing}, vol. 16.5, pp. 785--801, 2002.

\bibitem{Johan7}
J.~Schoukens, T.~Dobrowiecki, and R.~Pintelon, ``Parametric and nonparametric
  identification of linear systems in the presence of nonlinear distortions-a
  frequency domain approach,'' \emph{Automatic Control, IEEE Transactions on},
  vol.~43, no.~2, pp. 176--190, Feb 1998.

\bibitem{Johan3}
J.~Schoukens, R.~Pintelon, and Y.~Rolain, \emph{Mastering System Identification
  in 100 Exercises}, L.~Hanzo, Ed.\hskip 1em plus 0.5em minus 0.4em\relax John
  Wiley \& Sons, Inc., 2012.

\bibitem{Maarten1}
M.~Schoukens, K.~Tiels, M.~Ishteva, and J.~Schoukens, ``Identification of
  parallel wiener-hammerstein systems with decoupled static nonlinearity,''
  \emph{IFAC world congress}, 2014.

\bibitem{Johan6}
J.~Schoukens, J.~G. Nemeth, P.~Crama, Y.~Rolain, and R.~Pintelon, ``Fast
  approximate identification of nonlinear systems,'' \emph{Automatica}, vol.
  39(7), pp. 1267--1274, 2003.

\bibitem{Zumba}
H.~Zumbahlen, Ed., \emph{Linear Circuit Design Handbook}.\hskip 1em plus 0.5em
  minus 0.4em\relax Newnes, 2008.

\bibitem{Chen}
W.-K. Chen, Ed., \emph{The Circuits and Filters Handbook}.\hskip 1em plus 0.5em
  minus 0.4em\relax CRC Press, 2002.

\bibitem{Johan5}
J.~Schoukens, J.~Swevers, J.~Paduart, D.~Vaes, K.~Smolders, and R.~Pintelon,
  ``Initial estimate for block structure nonlinear systems with feedback,''
  \emph{Proc. Int. Symp. Nonlinear Theory Appl.}, pp. 622--625, 2005.

\bibitem{Istvan}
I.~Koll�r, R.~Pintelon, J.~Schoukens, and G.~Simon, ``Implementing a
  graphical user interface and automatic procedures for easier identification
  and modeling,'' \emph{IEEE Instrum. Meas. Magazine}, pp. 19--26, 2003.

\end{thebibliography}

\end{document}